\documentclass[A4paper]{article}

\usepackage{graphicx}
\usepackage{amsmath}
\usepackage{amssymb}
\usepackage{rotating}
\usepackage{multirow}
\usepackage{verbatim}
\usepackage{color}
\usepackage{subfigure}
\usepackage{arydshln}
\usepackage[toc,page]{appendix}
\usepackage{tikz}
\usetikzlibrary{bayesnet}
\usepackage{fullpage}

\newcommand{\balpha}{\boldsymbol{\alpha}}
\newcommand{\bx}{\boldsymbol{x}}

\newcommand{\bY}{\boldsymbol{Y}}

\newcommand{\bbeta}{\boldsymbol{\beta}}
\newcommand{\btheta}{\boldsymbol{\theta}}

\newcommand{\bs}{\boldsymbol{s}}
\newcommand{\ba}{\boldsymbol{a}}
\newcommand{\be}{\boldsymbol{e}}
\newcommand{\bdelta}{\boldsymbol{\delta}}
\newcommand{\bnu}{\boldsymbol{\nu}}
\newcommand{\btau}{\boldsymbol{\tau}}
\newcommand{\bz}{\boldsymbol{0}}

\title{Emulation of multivariate simulators using thin-plate splines with application to atmospheric dispersion}

\author{Veronica E. Bowman
\thanks{
Defence Science and Technology Laboratory, Salisbury, SP4 0JQ, UK
(vbowman@mail.dstl.gov.uk). 
}
\and
David C. Woods
\thanks{
Corresponding author. 
Southampton Statistical Sciences Research Institute, University of Southampton, Southampton, SO17 1BJ, UK 
(d.woods@southampton.ac.uk). 
}
}

\begin{document}
\maketitle

\begin{abstract}
It is often desirable to build a statistical emulator of a complex computer simulator in order to perform analysis which would otherwise be computationally infeasible.  We propose methodology to model multivariate output from a computer simulator taking into account output structure in the responses.  The utility of this approach is demonstrated by applying it to a chemical and biological hazard prediction model.  Predicting the hazard area which results from an accidental or deliberate chemical or biological release is imperative in civil and military planning and also in emergency response.   The hazard area resulting from such a release is highly structured in space and we therefore propose the use of a thin-plate spline to capture the spatial structure and fit a Gaussian process emulator to the coefficients of the resultant basis functions.  We compare and contrast four different techniques for emulating multivariate output: dimension-reduction using (i) a fully Bayesian approach with a principal component basis, (ii) a fully Bayesian approach with a thin-plate spline basis, assuming that the basis coefficients are independent, and (iii) a ``plug-in'' Bayesian approach with a thin-plate spline basis and a separable covariance structure; and (iv) a functional data modeling approach using a tensor-product (separable) Gaussian process.  We develop methodology for the two thin-plate spline emulators and demonstrate that these emulators significantly outperform the principal component emulator. Further, the separable thin-plate spline emulator, which accounts for the dependence between basis coefficients, provides substantially more realistic quantification of uncertainty, and is also computationally more tractable, allowing fast emulation. For high resolution output data, it also offers substantial predictive and computational advantages over the tensor-product Gaussian process emulator.\\[1ex]
This paper will appear in the Journal of Uncertainty Quantification. \\
\end{abstract}

%

\section{Introduction}
\label{intro}

The simulation of scientific and engineering systems via complex mathematical models has become a common method of gaining knowledge about processes where physical experimentation is infeasible or unaffordable. Encapsulated in computer codes or simulators, many of these models require substantial computing time to evaluate the response for a given set of inputs. For even moderately expensive simulators, the computational resources required to perform, for example, Monte Carlo inference may be prohibitive in practice. Hence, building an emulator or surrogate for the computer model trained on a, usually small, set of simulator evaluations has become standard practice; see for example, the seminal paper of Sacks et al. \cite{sacks}, Kennedy et al. \cite{kaco}, who presented a number of case studies of such computer experiments, and the book-length treatments of Santner et al. \cite{swn}, Fang et al. \cite{fls} and Forrester et al. \cite{fsk}. Computationally cheap emulators allow for real-time decision making and greater scientific understanding through, for example, sensitivity and uncertainty analyses.  

Statistical modelling is a common method for constructing emulators. Essentially, simulator output is treated as a realisation of a stochastic process, and regression models are fitted to the data in order to approximate the relationship between the simulator inputs and the outputs. The most common emulator is the Gaussian process (e.g. \cite{rasmussen}), a smooth non-parametric interpolator. An emulator based on a statistical model allows for prediction of the simulator at untested inputs and quantification of the associated uncertainty. For deterministic simulators, as considered in this paper, this uncertainty is a result of incomplete knowledge of the simulator across the whole input space and approximation error from the regression model.

Modern applications increasingly involve highly multivariate simulators, with each run of the simulator producing data from a curve, surface or other high-dimensional output structure. The standard approach is to perform dimension reduction on the multivariate output using a set of appropriate basis functions and then use scalar emulation methods, such as the Gaussian process, on the basis coefficients. We delay our discussion of the related statistical literature to Section~\ref{multiem}.

Motivated by a simulator of chemical and biological dispersion, the contribution of this paper is to propose the use of thin-plate splines as basis functions for multivariate data with spatial structure, and to develop the necessary methodology for their application. For a two-dimensional output, a thin-plate spline basis provides a spatial mapping using the proximity of data to a set of knots (see Section~\ref{tps}). Splines have also recently been employed for computer experiment modelling \cite{cmmkbgrsd} as an alternative to Gaussian process models. We implement both a fully Bayesian thin-plate spline emulator using Markov chain Monte Carlo and also a ``plug-in'' emulator (e.g. \cite{ko}) with a separable covariance structure \cite{rougier2008} and correlation parameters estimated using a validation data set. We provide a detailed comparison of these competing methodologies with the application of a functional data model using a Gaussian process defined on both the input and output domains. 

Dispersion simulators have widespread application in environmental monitoring, civilian emergency planning and military applications, for example in the protection against terrorism threats. Available simulators range from quite simple Gaussian plume models (e.g. \cite{clarke}), through Gaussian puff models (e.g. \cite{sykes}) to computationally expensive Lagrangian models (e.g. \cite{jones}).

In this paper, the problem of emulating a multivariate simulator built from a Gaussian puff model is considered. This model accounts for both the effects of an urban environment and the underlying terrain features. For each run of the simulator, the response of interest is the dosage, defined as the integrated concentration over time, measured at a large number of points in a two-dimensional geographical domain.   Inputs to the simulator include meteorological variables (such as wind speed and direction) and variables related to the source of the dispersion (such as location and size of release).   While relatively fast  ($\sim$ 1 minute per run for a complete spatial output grid) and simple to compute, the dispersion model is employed in the optimization of sensor placements. Therefore computationally inexpensive surrogates are required that can provide accurate high-resolution prediction on the spatial output grid. 

Previously, a number of authors have considered the univariate problem of \textit{calibrating} (relatively simple) dispersion models using actual dispersion data obtained under a single, but uncertain, set of meteorological and source conditions; see, for example, \cite{sf1993}, \cite{ko}, \cite{pr2004} and \cite{rrg2009}. Typically, the aim of such work is to solve the inverse problem of identifying unknown features of the source.  

The remainder of this paper is organised as follows. In Section~\ref{multiem}, we introduce and describe the multivariate emulator, dimension reduction techniques for multivariate outputs and separable covariance structures. The necessary methods for thin-plate spline emulation are developed in Section~\ref{tps}, and applied to an illustrative dispersion model in Section~\ref{app}. In that section, we also compare our methodology to applying dimension reduction via principal components and to a tensor-product, separable Gaussian process model. We explore the effectiveness of the methodology for high-resolution prediction. In Section~\ref{artex}, we compare dimension-reduction and functional data modelling on an artificial environmental example. We conclude in Section~\ref{disc} with some discussion and areas for future research. The appendices provide mathematical and computational details of the methods.


\section{Multivariate emulation}
\label{multiem}

%
%
%
%
%

Let $\bx_i = (x_{1i},\ldots,x_{di})^{\rm T}\in\mathcal{X}\subseteq\mathbb{R}^d$ be the vector of input values at which the $i$th run of the simulator is performed, i.e. the $i$th input point ($i=1,\ldots,n$), and let $\bY(\bx_i) = \left[Y(\bx_i,\bs_1),\ldots,Y(\bx_i,\bs_r)\right]^{\rm T}$ be the vectorised output from this run. The vector $\bs_j=(s_{1j},\ldots,s_{qj})^{\rm T}\in\mathcal{S}$ locates the $j$th output in the $q$ dimensional output domain $\mathcal{S}\subseteq\mathbb{R}^q$ ($j=1,\ldots,r$). For example, for a two-dimensional simulator, $q=2$ and $\bs_j=(s_{1j},s_{2j})^{\rm T}$ which may be the geographical coordinates of the response. We denote the complete output grid as an $r\times q$ matrix $S = (\bs_1, \ldots, \bs_r)^{\rm T}$.

We achieve dimension reduction through the assumption of a linear model for the response:

\begin{equation}\label{mvreg}
\bY(\bx_i) = \sum_{k=1}^p\ba_k(S)\beta_k(\bx_i) + \be_i\,.
\end{equation}

\begin{figure}
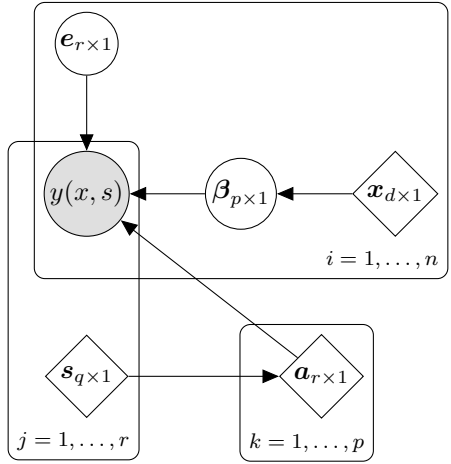

      \centering
      \tikz{ %
      	\node[latent] (e) {$\be_{r\times 1}$} ; %
      	\node[obs, below=of e] (y) {$y(x,s)$} ; %
	\node[latent, right=1cm of y] (beta) {$\bbeta_{p\times 1}$} ; %
        \node[det, right=1cm of beta] (x) {$\bx_{d\times 1}$} ; %
        \node[det, below=1.3cm of y] (s) {$\bs_{q\times 1}$} ; %
        \node[det, right=2cm of s] (a) {$\ba_{r\times 1}$} ; %
	\plate {plate1}{(e) (y) (beta) (x)} {$i=1,\ldots,n$} ; %
	\plate {plate2}{(s) (y)} {$j=1,\ldots,r$} ; %
	\plate{plate3}{(a)} {$k=1,\ldots,p$} ; %
	\edge {e} {y} ; %
	\edge {beta} {y} ; %
	\edge {x} {beta} ; %
	\edge {s} {a} ; %
	\edge {a} {y} ; %
         }
	\caption{\label{fig:gm}Representation of model~\eqref{mvreg} as a directed graph with edges representing dependencies, the shaded node representing the observable output, unshaded circular nodes representing unobservable features of the model, and diamond nodes representing deterministic model inputs. Subscripts indicate size of vectors and plates indicate the sizes of data sets.}
    \end{figure}

\noindent In~(\ref{mvreg}), $\ba_1(S),\ldots,\ba_p(S)$ are a set of $r\times 1$ basis vectors which are assumed independent of $\bx$ but which may depend on the output grid $S$, the coefficients $\bbeta(\bx) = \left[\beta_1(\bx),\ldots,\beta_p(\bx)\right]^{\rm T}$ are functions of the input variables $\bx$ and $\be_i$ is a $r$-vector of errors resulting from the basis function approximation. Figure~\ref{fig:gm} summarizes this hierarchical model and demonstrates the dependence of $\bbeta(\bx_1),\ldots,\bbeta(\bx_n)$ on $\bx_1,\ldots,\bx_n$, and $\ba_1,\ldots,\ba_p$ on $\bs_1,\ldots,\bs_r$. 

In our motivating application, interest is only in prediction at the $\bs_j$, and hence for each output location we mean-center and standardize the data $\left[Y(\bx_1, \bs_j), \ldots, Y(\bx_n, \bs_j)\right]$ to have variance equal to one. We complete the hierarchical specification of the model through choice of prior distributions. For $\beta_k(\bx)$, we assume the impact of the input variables is modeled using a Gaussian process 

\begin{equation}\label{betaprior}
\beta_k(\bx) | \tau_k, \btheta_k  \sim \mbox{GP}\left(0, \tau_k\rho(\bx, \bx^\prime;\,\btheta_k)\right)
\end{equation}

\noindent where $\mbox{GP}$ denotes a Gaussian process and $\rho(\bx, \bx^\prime;\,\btheta)$ is a correlation function dependent on input vectors $\bx$ and $\bx^\prime$. Scale and correlation length hyper-parameters are labeled $\tau_k$ and $\btheta_k = (\theta_{1k}, \ldots, \theta_{dk})^{\rm T}$ respectively. We discuss the choice of joint distribution of $\beta_k(\bx)$ and $\beta_{k^\prime}(\bx^\prime)$ below. For $\be_i$ we assume

\begin{equation}\label{errorprior}
\be_i | \sigma^2 \sim N(\boldsymbol{0}_{r},I_{r}\sigma^2)\,,
\end{equation}

\noindent with  $I_{r}$ the $r\times r$ identity matrix and $\sigma^2 > 0$. The choice of a independent multivariate prior distribution for $\be_i$ assumes that the regression function $\sum_{k=1}^p\ba_k(S)\beta_k(\bx_i)$ is detailed enough to capture the vast majority of the variation in the data. The adequacy of this assumption for a particular application can be checked via standard residual diagnostics.


%
%
%

To specify the joint distribution of $\beta_k(\bx)$ and $\beta_{k^\prime}(\bx^\prime)$, $k, k^\prime = 1,\ldots, p$, we consider two simplifying cases.

\begin{enumerate}
\item[(i)] Independence of $\beta_k(\bx)$ and $\beta_{k^\prime}(\bx^\prime)$ for $k\ne k^\prime$, that is 
$$\mbox{Cov}\left\{\beta_k(\bx), \beta_{k^\prime}(\bx^\prime)\right\} = \tau_k\rho(\bx,\bx^\prime;\,\btheta_k)\mathbb{I}(k=k^\prime)\,,$$
with $\mathbb{I}$ the indicator function. Let $\bbeta_k = (\beta_k(\bx_1),\ldots,\beta_k(\bx_n))^{\rm T}$ and $\bbeta = (\bbeta_1^{\rm T},\ldots,\bbeta_p^{\rm T})^{\rm T}$. Also, let $\tau_kW_k$ denote the $n\times n$ variance-covariance matrix of $\bbeta_k$, with $ij$th entry $\tau_k\rho(\bx_i,\bx_j;\,\btheta_k)$. Then, the block diagonal variance-covariance matrix of $\bbeta$ is given by 
\begin{equation}\label{eq:indep}
\mbox{Cov}\left\{\bbeta\right\} = \mbox{diag}\left(\tau_1W_1, \ldots, \tau_pW_p\right)\,.
\end{equation}

\item[(ii)] A product covariance structure, with
$$\mbox{Cov}\left\{\beta_k(\bx), \beta_{k^\prime}(\bx^\prime)\right\} = \tau\rho(\bx,\bx^\prime;\,\btheta)\times\phi(\bs, \bs^\prime;\,\bnu)\,,$$
with $\phi(\bs, \bs^\prime;\,\bnu)$ a function depending on output locations $\bs$ and $\bs^\prime$ and parameters $\bnu = (\nu_1, \ldots. \nu_q)^{\rm T}$. Now, the variance-covariance matrix of $\bbeta$ has the form
\begin{equation}\label{eq:tpc}
\mbox{Var-Cov}\left\{\bbeta\right\} = \tau W \otimes V\,,
\end{equation}
with $\tau W$ being the $n \times n$ common variance-covariance matrix for $\bbeta_k$ with $ij$th entry $\tau\rho(\bx_i,\bx_j;\,\btheta)$ and $V$ a $p \times p$ scale matrix for $\bbeta(\bx_i) = \left[\bbeta_1(\bx_i), \ldots, \bbeta_p(\bx_i)\right]^{\rm T}$ with entries defined via $\phi$. The exact form of $\phi$ and construction of $V$ depends on the basis chosen; we discuss the choice of $V$ for our thin-plate spline emulators in Section~\ref{tps}.
\end{enumerate}

Point prediction of the response at a untried input point, $\bx^\star$, is via

\begin{equation}\label{mvregpred}
\hat{\bY}(\bx^\star) = \sum_{k=1}^p\ba_k(S)\hat{\beta}_k(\bx^\star)\,,
\end{equation}

\noindent where $\hat{\beta}_k(\bx^\star)$ is an appropriate summary of the posterior distribution of $\beta_k(\bx^\star)$. There is no conjugate prior distribution available for unknown $\btheta$ and so to obtain the full marginal distribution, numerical methods must be used, such as Markov chain Monte Carlo (MCMC; \cite{of}, ch. 10) with a sample from the posterior predictive distribution obtained by plugging samples from $\beta_k(\boldsymbol{x}^\star)|Y$ into~\eqref{mvregpred}. Alternatively, a maximum a posteriori (MAP), or other estimator, of $\btheta$ can be substituted into the conditional posterior distribution of $\beta_k(\bx^*)$.

Similar models have been developed and applied by a variety of authors. Campbell et al. \cite{cmw} considered a variety of basis functions for one-dimensional functional responses, including orthogonal polynomials and data-adaptive choices such as principal components and partial least squares. A wavelet basis was used by Bayarri et al. \cite{bbcglpppsw} in a calibration problem with a functional response. 
Higdon et al. \cite{hgwr} used a principal components basis for an example of cylinder deformation from the Manhattan project. These latter authors had a two-dimensional output grid, consisting of angles and times, and were again concerned with model calibration.

Principal component analysis (PCA; \cite{jolliffe}) is a dimension-reduction technique that has been commonly used in the literature for defining a new, orthogonal basis for a set of multivariate data. For a $r\times n$ matrix $Y=[\bY(\bx_1),\ldots,\bY(\bx_n)]$, the principal components are defined through the singular value decomposition of $Y = \Lambda\Sigma\Omega^{\rm T}$, where $\Lambda$ is a $r\times n$ orthogonal matrix whose columns are the left singular vectors of $Y$, $\Sigma$ is a $n\times n$ diagonal matrix holding the singular values of $Y$ and $\Omega$ is an $n\times n$ orthonormal matrix whose columns are the right singular vectors. A $p$-dimensional principal component basis $\ba_1,\ldots,\ba_p$ is given by the first $p$ columns of $n^{-1/2}\Lambda\Sigma$, with corresponding weights $\beta_k(\bx_i)$ given by entry $(k,i)$ in $n^{1/2}\Omega$ $(k = 1,\ldots, p;\, i=1,\ldots,n)$.  The basis functions are orthogonal and have the property of defining subspaces with the largest variance, see \cite{htf}, ch.3. The data-dependent nature of the principal components provides a flexible non-parametric modelling approach. However, it does not take the structure of the output grid $S$ into account.

\section{Thin-plate regression splines}
\label{tps}

To provide a set of flexible and data-driven basis vectors $\ba_1(S),\ldots,\ba_p(S)$ that maintain the multidimensional spatial structure inherent in $S$, we use thin-plate regression splines (TPRS) \cite{wood2003}. To define the TPRS basis, we start with the $r\times r$ matrix $E$ having $uv$th entry $\eta_{lq}(||\bs_u-\bs_v||)$, with $||\cdot||$ defined as Euclidean distance and

$$
\eta_{lq}(t)=\left\{
\begin{array}{cc}
\frac{(-1)^{l+1+q/2}}{2^{2l-1}\pi^{q/2}(l-1)!(l-q/2)!}t^{2l-q}\log(t) & \mbox{for } q \mbox{ even}\,,\\
& \\
\frac{\Gamma(q/2-l)}{2^{2l}\pi^{q/2}(l-1)!}t^{2l-q} & \mbox{for } q \mbox{ odd}\,,
\end{array}
\right.
$$

\noindent with $2l>d$ controlling the smoothness of the spline ($l=2$ corresponds to a smoothness penalty in terms of second derivatives of the response function, and is adopted in this paper). A thin-plate spline for the $i$th simulator run is then the solution to

\begin{equation}\label{tpseq}
\mbox{minimize} \qquad ||\bY(\bx_i) - E\bdelta_i - T\balpha_i||^2 + \lambda_i\bdelta_i^{\rm T}E\bdelta_i \qquad \mbox{subject to } T^{\rm T}\bdelta_i=\boldsymbol{0}\,,
\end{equation} 

\noindent with respect to $\bdelta_i$ and $\balpha_i$ for $\lambda_i\ge 0$ $(i=1,\ldots,n)$. The matrix $T$ holds basis vectors corresponding to orthogonal polynomials in $\mathbb{R}^{q}$ of degree less than $l$. In our applications with $l=2$ and a regular lattice output grid, $T$ is an $r\times 3$ matrix with $j$th row $[1 \mid \bs_j^{\rm T}]$. The thin-plate spline is therefore the solution to a penalised least squares problem.

To avoid problems of choosing ``knot locations'' (e.g. a subsample of the output grid) in selecting a regression basis, Wood \cite{wood2003} defined a TPRS basis as a rank $m$ approximation to the spline, with basis vectors given by the columns of $U_{m}D_{m}Z_{m}$ and $T$, where $D_{m}$ is an $m\times m$ diagonal matrix holding the $m$ largest eigenvalues of $E$ ordered by absolute value, $U_{m}$ is an $r \times m$ matrix holding the corresponding eigenvectors and $Z_{m}$ is an $m$ dimensional orthogonal column basis such that $T^{\rm T}U_{m}Z_{m}=\boldsymbol{0}$. That is, an approximation to problem~(\ref{tpseq}) with $\lambda_i=0$ (i.e. unpenalised) is given by

\begin{equation}\label{tpseq2}
\mbox{minimize} \qquad ||\bY(\bx_i) - U_{m}D_{m}Z_{m}\delta_i^\prime - T\balpha_i||^2\,,
\end{equation} 

\noindent with respect to $\bdelta_i^\prime$ and $\balpha_i$, where $\bdelta_i^\prime = Z_m^{\rm T}U_m^{\rm T}\bdelta_i$ and condition $T^{\rm T}U_{s}Z_{s}=\boldsymbol{0}$ ensures $T^{\rm T}\bdelta_i=\boldsymbol{0}$.


Hence, we set the first $m$ basis vectors in~\eqref{mvreg}, $\ba_1(S),\ldots,\ba_{m}(S)$, to be the columns of $U_{m}D_{m}Z_{m}$, and the second $p-m$ basis vectors, $\ba_{m+1}(S),\ldots,\ba_{p}(S)$, to be the columns of $T$. For $i=1,\ldots,n$, we set 
$$
\bbeta(\bx_i) = \left[\bbeta_{m}(\bx_i)^{\rm T},\bbeta_{p-m}(\bx_i)^{\rm T}\right]^{\rm T} = \left[\left(\bdelta_i^{\prime\star}\right)^{\rm T}, \left(\balpha^\star\right)^{\rm T}\right]^{\rm T}\,,
$$
the solution to~\eqref{tpseq2}. We require a constant scale matrix for $\bbeta(\bx_i)$ for all $\bx_i$ in order to construct the separable variance-covariance matrix~\eqref{eq:tpc} for the complete parameter vector $\bbeta$ and to allow prediction for untried input points. Hence, rather than derive the variance-covariance matrix for $\bbeta(\bx_i)$ conditional on $\bY(\bx_i)$, we assume $\mbox{Cov}\left(\bdelta_i^\star\right) \propto Q$ where $\bdelta_i^\star = U_mZ_m\bdelta_i^{\prime\star}$ and $Q$ has $uv$th entry given by $\rho(\bs_u,\bs_v;\,\bnu)$ $(u,v=1,\ldots,r)$, a spatial correlation function defined on the output locations. It then follows that $\mbox{Cov}\{\bbeta_{m}(\bx_i)\} \propto Z^{\rm T}_{m}U^{\rm T}_{m}QU_{m}Z_{m}$. Further, we assume $\bbeta_{m}(\bx_i)$ and $\bbeta_{p-m}(\bx_i)$ are independent, with $\mbox{Cov}\{\bbeta_{p-m}(\bx_i)\}\propto I_{p-m}$, leading to scale matrix
$$
V = \left(
\begin{array}{cc}
Z^{\rm T}_{m}U^{\rm T}_{m}QU_{m}Z_{m} & \boldsymbol{0}_{m\times p} \\
\boldsymbol{0}_{m\times p}^{\rm T} & I_{p-m}
\end{array}
\right)\,.
$$ 


\section{Application to the dispersion simulator}
\label{app}
In this section, the methodology from Sections~\ref{multiem} and~\ref{tps} is applied to an exemplar dispersion simulator. As described in Section~\ref{intro}, the response of interest is dosage (integrated concentration over time). The resultant dosage map is known as a hazard prediction and forms the basis for many modelling systems designed to mitigate hazard such as source-term estimators \cite{rrg2009}, sensor placement optimizers and training scenarios. The motivating system for this paper is a sensor placement tool which requires rapid hazard predictions over which the probability of detection of a chemical or biological release is numerically maximized via the placement of a set of sensors. For other applications, the methodology could be applied to emulate concentration at a given time or time could be included as an extra dimension to either the input or output (cf \cite{co2010}). 

In the application of interest, it is key that (i) a hazard prediction can be made rapidly, to assess performance of a sensor grid against a particular threat in real time, and (ii) there is accurate prediction of the dosage on a common high-resolution spatial output grid for untried values of the input variables.  Our data comes from the Urban Dispersion Model (UDM), an example of a Gaussian puff model. In these simulators, continuous releases are modelled as a serious of instantaneous releases (``puffs'') with Gaussian profiles in the ``$x$'' and ``$y$'' planes (forming a bell shape). The UDM can model dispersion across complex terrain such as urban areas by including the effect of buildings, in addition to underlying terrain (for example, hills or valleys), by tracking the interaction of individual puffs with local geographical features.  

\begin{figure}[!t]
\centering
\subfigure[Perspective Plot]{
\includegraphics[scale=0.5]{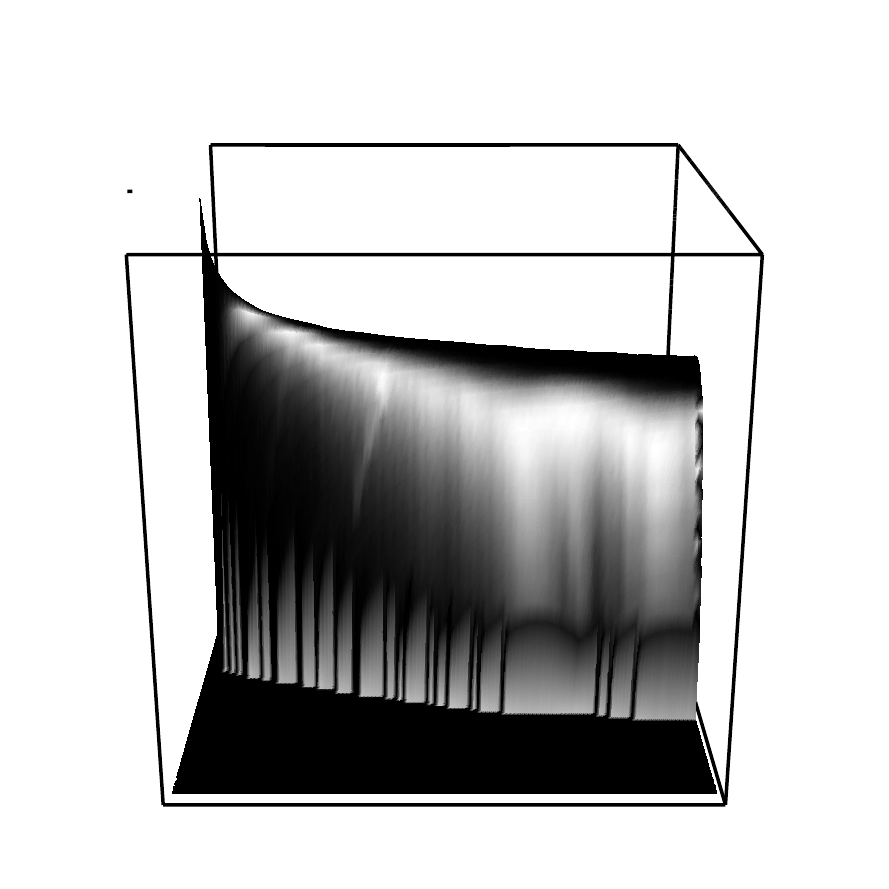}
}
\subfigure[Contour Plot]{
\includegraphics[scale=0.25]{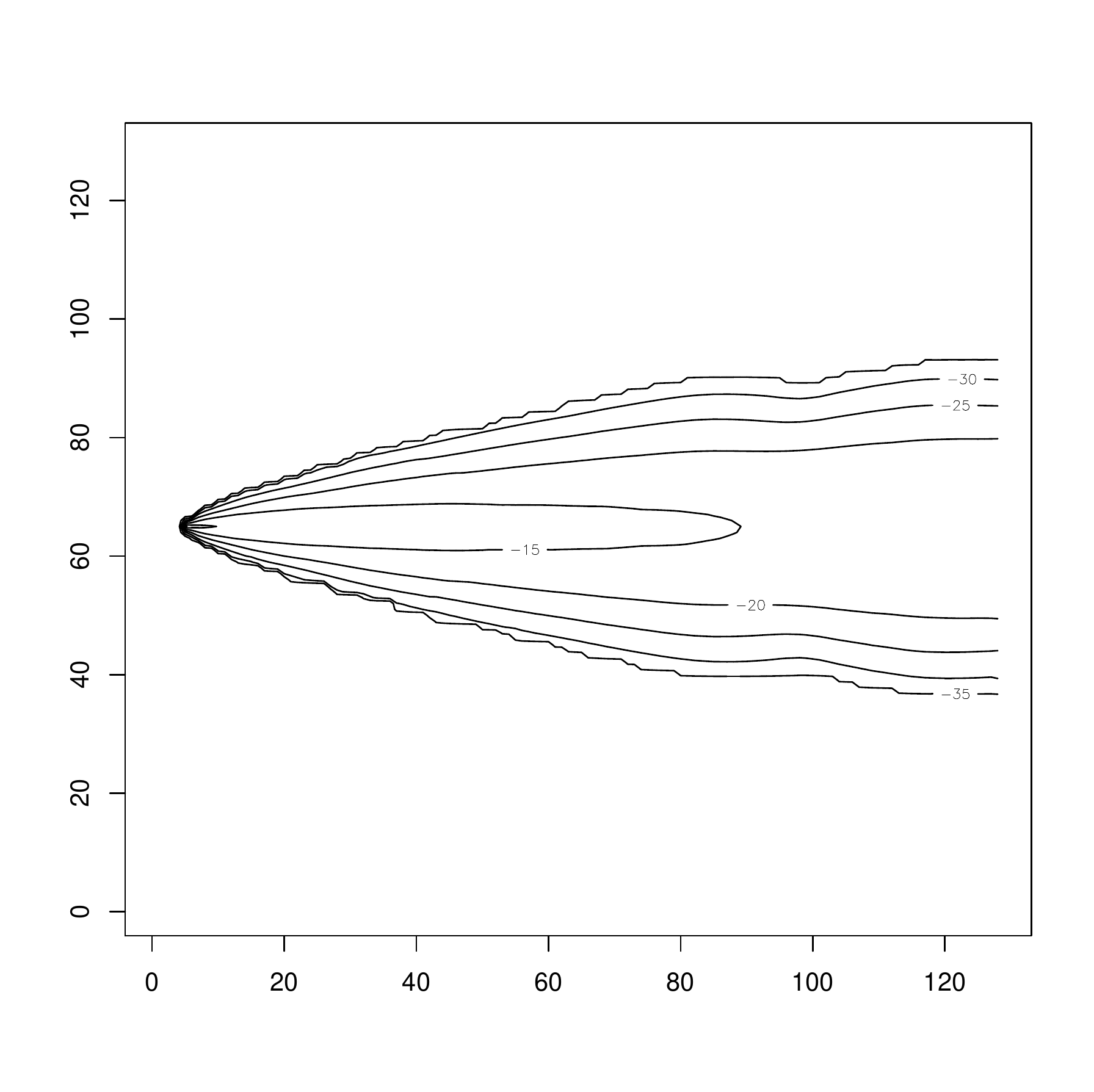}
\label{fig:actcontour}
}
\label{fig:actualPlume}
\caption{\label{fig:persp}A typical dosage output from the dispersion model output on a logged scale}
\end{figure}

There are numerous inputs to a dispersion simulator, including those related to the release (e.g. \cite{bw2012a}). For this application, the input space is limited to the $d=2$ dimensions which, from previous experience, are known to have greatest effect on dosage: mass of the release ($0\le x_1 \le 50$kg) and wind speed ($5 \le x_2 \le 30$m/s). For a given terrain, other variables known to affect dosage, such as wind direction and location of release, can be specified post-simulation through rotation and translation operations. Typical simulator responses exhibit a very rapid drop in dosage upwind of the release and a complex downwind gradual reduction in dosage. There is also different behavior in the down-wind (``$x$'') and cross-wind (``$y$'') directions. This behavior is not well described by standard exponential functions; see Figure~\ref{fig:persp}. 

Our data results from simulator runs on a ``uniform'' urban area representative of a medium-rise city environment outside of the central business district. Accurate high-resolution spatial prediction is important for both comparing competing sensor placements in the optimization tool and ensuring assessments of chosen sensor placements are realistic summaries of actual system performance. Errors of only a few tens of meters in sensor placement, perhaps caused by small prediction errors in the modeling, can result in large discrepancies in predicted dosage and therefore lead to substantial underperformance of a sensor system against a specific threat.  Any emulator must therefore be capable of producing rapid predictions on a high-resolution spatial grid and even small improvements in predictive performance are practically valuable.

For the $i$th run, dosage is output on a $k\times k$ grid ($i=1,\ldots,n$). Hence, $\bs_j=(s_{1j}, s_{2j})^{\rm T}$ locates the $j$th (standardised) geographical position at which dosage is calculated $(j=1,\ldots,r)$. The output grid is common to all runs; if this were not the case, linear interpolation could be used to map the outputs from different simulator runs to a common output grid.  The vector $\bY(\bx_i)$ holds the $r=k^2$ dosages for the $i$th run; we in fact model $\log\left(\bY(\bx_i)+1\right)$. The simulator input values are held in $\bx_i=(x_{1i},x_{2i})^{\rm T}$ $(i = 1, \ldots, n)$.   

   
We apply and compare four emulation approaches:
\begin{enumerate}
\item A fully Bayesian approach using a principal components basis (PC-GP emulator).
\item A fully Bayesian approach using a TPRS basis and assuming independence of the elements of $\bbeta(\bx_i)$, cf~\eqref{eq:indep} (iTPRS-GP emulator).
\item An empirical Bayes approach using a TPRS basis and assuming a separable covariance structure for $\bbeta$, cf~\eqref{eq:tpc} (sTPRS-GP emulator).
\item A empirical Bayes approach using a Gaussian process with a separable covariance structure defined on $\mathcal{X}\times\mathcal{S}$ (sGP emulator). 
\end{enumerate}

For emulators 1 and 2, the posterior predictive distribution for $\bY$ is obtained using MCMC with convergence assessed graphically using diagnostic plots. For these emulators, we also assume $V=I_p$; that is, the elements of $\bbeta(\bx_i)$ are assumed independent. While for emulator 1 this assumption has a heuristic justification via adoption of an orthogonal PC basis, we acknowledge that for emulator 2 it may lead to overconfident prediction intervals and an over-estimate of the effective sample size. However, it substantially reduces the computational burden of the MCMC algorithm and provides a benchmark for emulators 3 and 4. Here, the correlation length parameters $\btheta$ and $\bnu$ are chosen using validation simulation runs and substituted into the conditional posterior predictive density (see also \cite{rougier2008} and \cite{rgmr}).

Emulator 4 models the simulator output directly as a function of both $\bx$ and $\bs$, that is
$$
Y(\bx, \bs) \mid \tau, \btheta, \bnu \sim \mbox{GP}\left[0, \tau\rho(\bx, \bx^\prime;\,\btheta)\times\rho(\bs, \bs^\prime;\,\bnu)\right]\,.
$$
The assumption of a separable correlation structure results in a tensor-product variance-covariance matrix for $\bY = \left[\bY(\bx_1), \ldots, \bY(\bx_n)\right]$. See \cite{rougier2008} for a detailed discussion of this model, and extensions.

\subsection{Choice of correlation function}\label{corrfunsec}
In this paper, we use a squared exponential correlation function \cite{hgwr}, with parameterization 

\begin{equation}\label{corrfun}
\rho(\bx,\bx^\prime;\,\btheta) = \prod_{j=1}^{d}\theta_j^{4(x_{j}-x_{j}^\prime)^2}\,.
\end{equation}

\noindent  Assuming a standardized wind direction, which can be adjusted post-simulation, the main variation in the response will be axially aligned, making~\eqref{corrfun} a reasonable choice. A fully Bayesian approach requires a prior distribution for each $\theta_j$ and we assume common Beta prior densities with $\pi(\theta_k)\propto \theta_k^{a_\theta-1}(1-\theta_k)^{b_\theta-1}$. A nugget term was added to improve the conditioning of the variance-covariance matrix of~$\bbeta$. Clearly, alternative correlation functions could be employed; we also tried the Mat{\'e}rn correlation function but found for this application it led to less accurate results and did not improve numerical stability of the modeling. The addition of a nugget is a pragmatic method for reducing the sensitivity of the modelling results to underlying assumptions, including the choice of covariance function \cite{gl2012}.

\subsection{Design of the simulator runs}
\label{design}

The emulators were trained using $n=80$ simulator runs generated as a maximin Latin Hypercube sample \cite{mm}.  A further 75 simulator runs were generated randomly as a Monte Carlo sample, with 40 of these runs used for validation and choice of some prior hyper-parameters. The remaining 35 runs were used as test cases to assess emulator accuracy. Initially, all simulator runs were generated for a low-resolution $k=8$ regular lattice in $\mathcal{S}$ with no boundary points. We assess the more promising emulators on a high-resolution output grid in Section~\ref{dense}.

\begin{figure}[!t]
\begin{center}
\includegraphics[scale=.6,clip=true]{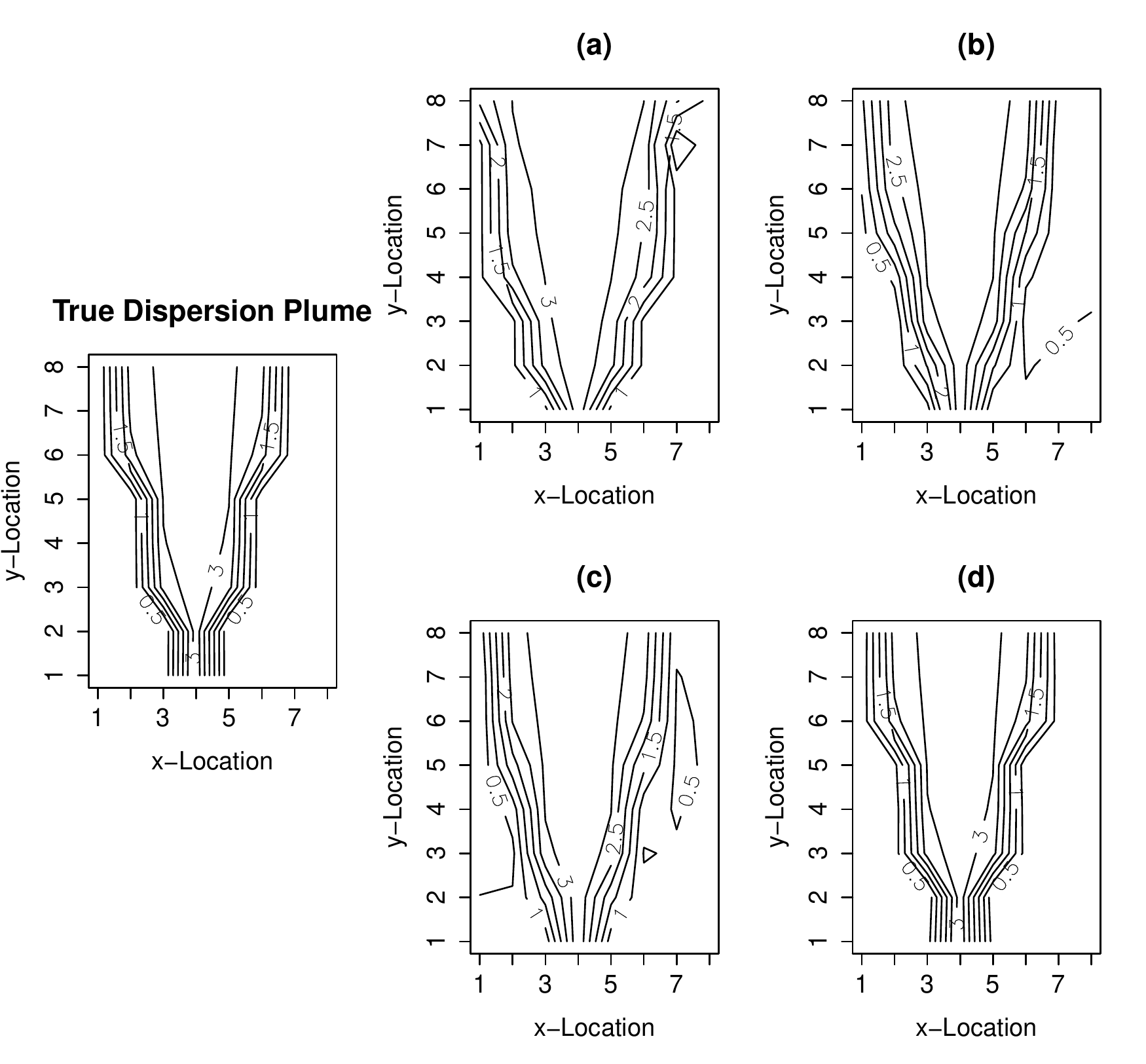}
\end{center}
\caption{\label{predict}{Actual and predicted log dosage for one test run for the (a) PC-GP, (b) iTPRS, (c) sTPRS and (d) sGP emulators.}}
\end{figure}

\subsection{Model fitting and choice of hyper-parameters}\label{fitting}

Bayesian inference for the PC-GP and iTPRS-GP emulators proceeded using MCMC and the likelihoods derived by Higdon et al. \cite{hgwr} (for PC-GP) and in Appendix~\ref{itpsappendix} (for iTPRS-GP). To reduce computational complexity for the iTPRS-GP emulator, we applied the Woodbury matrix identity (e.g. \cite{graybill}, Ch.8) to update the inverse and determinant of the expanded model matrix within the MCMC iterations.  This replaces the inversion of a $np \times np$ square matrix with the inversion of a $n \times n$ matrix when sampling each of the nugget, $\tau_k$ and $\btheta_k$. An outline of this implementation is given in Appendix~\ref{secwood}. The MCMC was run for 10000 iterations with a burn-in of 1000 which produced acceptable trace plots with no bias from starting location.  No thinning is performed as the trace plots showed the chains mixing well. Specification of the prior hyper-parameters for the common Gamma distributions for each $\tau^{-1}_k$ is simplified by the standardization of the simulator output, which leads to an expectation of the variance of each $\beta_k(\bx)$ being close to one. Hence, we chose an informative Gamma$(5,0.2)$ prior distribution, with density $f(\tau^{-1}) \propto \tau^{-4}\exp\left(-5/\tau\right)$, again following \cite{hgwr}. 

Other hyper-parameters for all four emulators, including the numbers of regression basis functions in~\eqref{mvreg}, were chosen to minimise the root mean squared error (RMSE) of prediction on the validation runs. Alternatives include choosing the hyper-parameters to maximise the integrated likelihood but use of the RMSE is consistent with our goal of accurate prediction for untried $\bx$. For all the dimension-reduced emulators (PC-GP, iTPRS-GP and sTPRS-GP), the number of basis functions $p$ was the main determinant of RMSE. The choice of $p$ controls the trade-off between detailed modeling of the training data and the generalization to untried cases. Increasing $p$ also increases the computational expense of the model fitting. For all three dimension-reduced emulators, we found relatively small values of $p$ minimized the RMSE.

For the PC-GP and iTPRS emulators, these hyper-parameters consisted of $a_\theta$ and $b_\theta$, common to the independent Beta prior distributions assumed for the entries of $\btheta_k$ (with density $f(\theta) \propto \theta^{a_\theta-1}(1-\theta)^{b_\theta-1}$), $a_\sigma$ and $b_\sigma$ for the Gamma distribution assumed for $\sigma^{-2}$, and the numbers of principal components (PC-GP) and thin plate spline basis functions (iTPRS-GP). We choose $a_\sigma=2$ to define a Gamma distribution with $\lim_{z\to 0^+}f(z) = 0$ and $\lim_{z\to 0^+}f^\prime(z)=b_\sigma^{-2}$, so that larger values of $b_\sigma$ result in higher prior density for larger values of $\sigma^{-2}$. In general, RMSE was lowest for lower values of $b_\sigma$, and also smaller numbers of basis functions. Various shapes of Beta prior density were tested and the results are fairly robust to the choices of $a_\theta$ and $b_\theta$. For both emulators, we chose $b_\sigma = 0.01$ and $a_\theta = 1$; for the PC-GP emulator, $p = 3$ and $b_\theta = 3$, and for the iTPRS-GP emulator, $p = 5$ and $b_\theta = 0.1$. Several similar hyper-parameter settings produced comparable RMSE.


For the sTPRS-GP and sGP emulators, correlation parameters $\btheta=(\theta_1,\theta_2)^{\rm T}$ and $\bnu=(\nu_1,\nu_2)^{\rm T}$ are chosen using the validation runs, together with hyper-parameters $a_\tau$ and $b_\tau$ for the Gamma prior distribution on $\tau$. The numerical results indicated that the RMSE was robust to the choice of $a_\tau$ and $b_\tau$, and a Gamma(1,1) prior distribution was chosen, giving near linearly decreasing support between 0 and 1. For the sTPRS-GP emulator, the minimum RMSE was achieved with $p=5$ and $\btheta=\bnu=(0.05,0.05)^{\rm T}$. For the sGP emulator, the minimum RMSE occurred with $\btheta = (0.5, 0.65)^{\rm T}$ and $\bnu = (0.8, 0.4)^{\rm T}$. For these two emulators, $\sigma^2$ is estimated from the mean squared error on the test data.

\begin{figure}[!t]
\begin{center}
\includegraphics[scale=.7,clip=true]{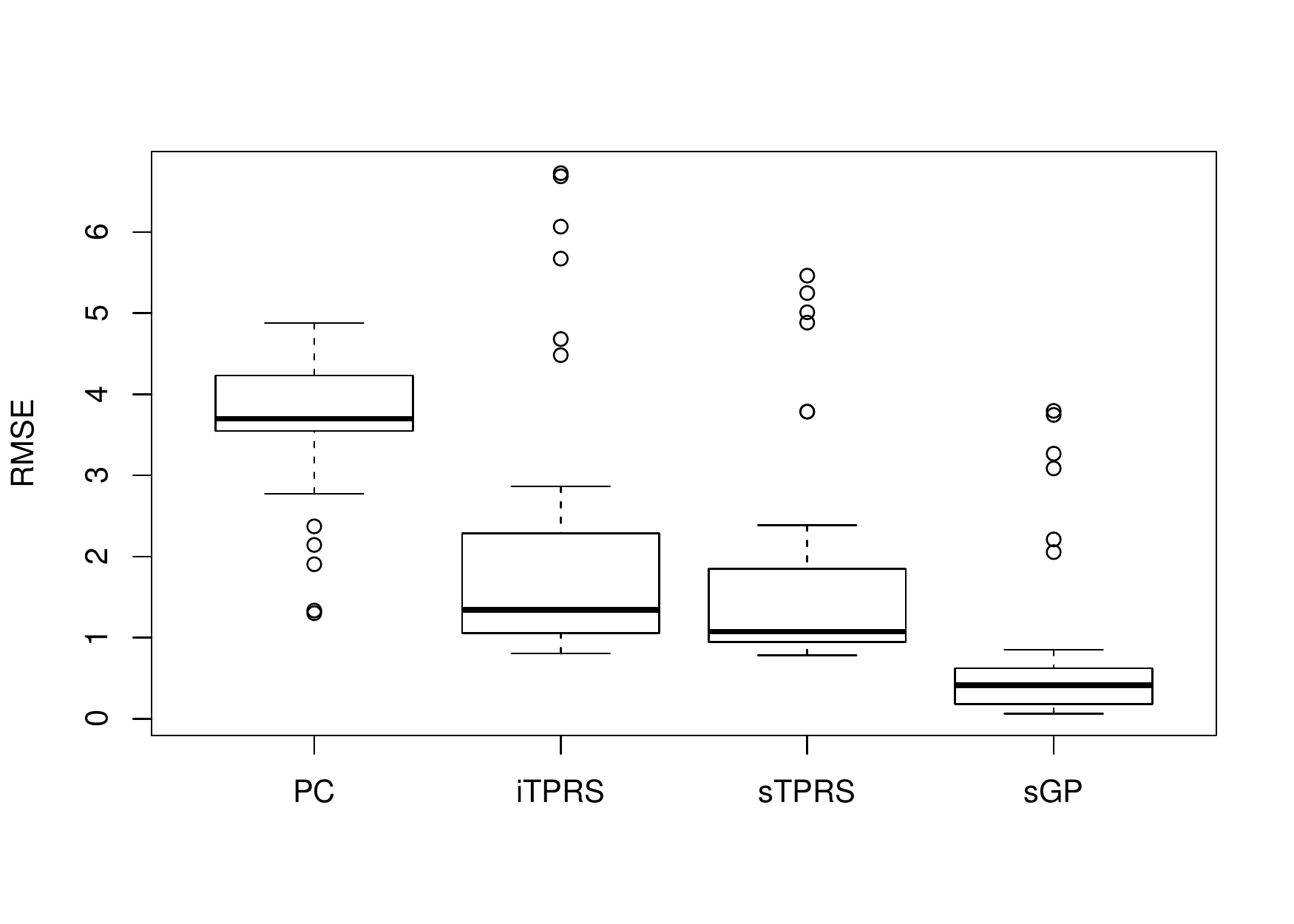}\\[-2ex]
\end{center}
\caption{\label{boxplot}Root mean squared error for each of the PC-GP, iTPRS-GP, sTPRS-GP and sGP emulators, calculated using the posterior predictive mean across the test set.}
\end{figure}
\subsection{Comparison of Emulators}\label{comp}

Each of the four emulators were used to predict the logged dosage for the 35 test runs using the posterior predictive mean.  All four emulators produced similar spatial dispersion to the true simulator, illustrated in Figure~\ref{predict}.  The PC-GP emulator has generally higher RMSE across all of the test runs; see Figure~\ref{boxplot}. The median RMSE for the PC-GP emulator (3.70) is considerably greater than the upper quartile of the iTPRS-GP (2.28) or sTPRS-GP (1.85) emulators. The sTPRS-GP emulator performs slightly better than the iTPRS-GP emulator, with lower quartiles of 1.07 for sTPRS-GP and 1.35 for iTPRS-GP. These differences demonstrate the advantage of modeling the within-run correlation between basis functions.  For this low-resolution spatial grid, the sGP emulator shows better performance than any of the other emulators, with lower and upper quartiles of 0.19 and 0.61 respectively. In Section~\ref{dense}, we investigate if the sGP emulator maintains this advantage over the sTPRS-GP emulator for a high-resolution output grid where the sGP emulator can only be trained on a subset of the output data.

\subsection{Uncertainty quantification}\label{uncertaintysec}

\newcommand{\plotscale}{0.35} 
\begin{figure}[!t]
\begin{center}
\begin{tabular}{cc}
(a) & (b) \\[-14ex]
\includegraphics[scale=\plotscale,clip=true]{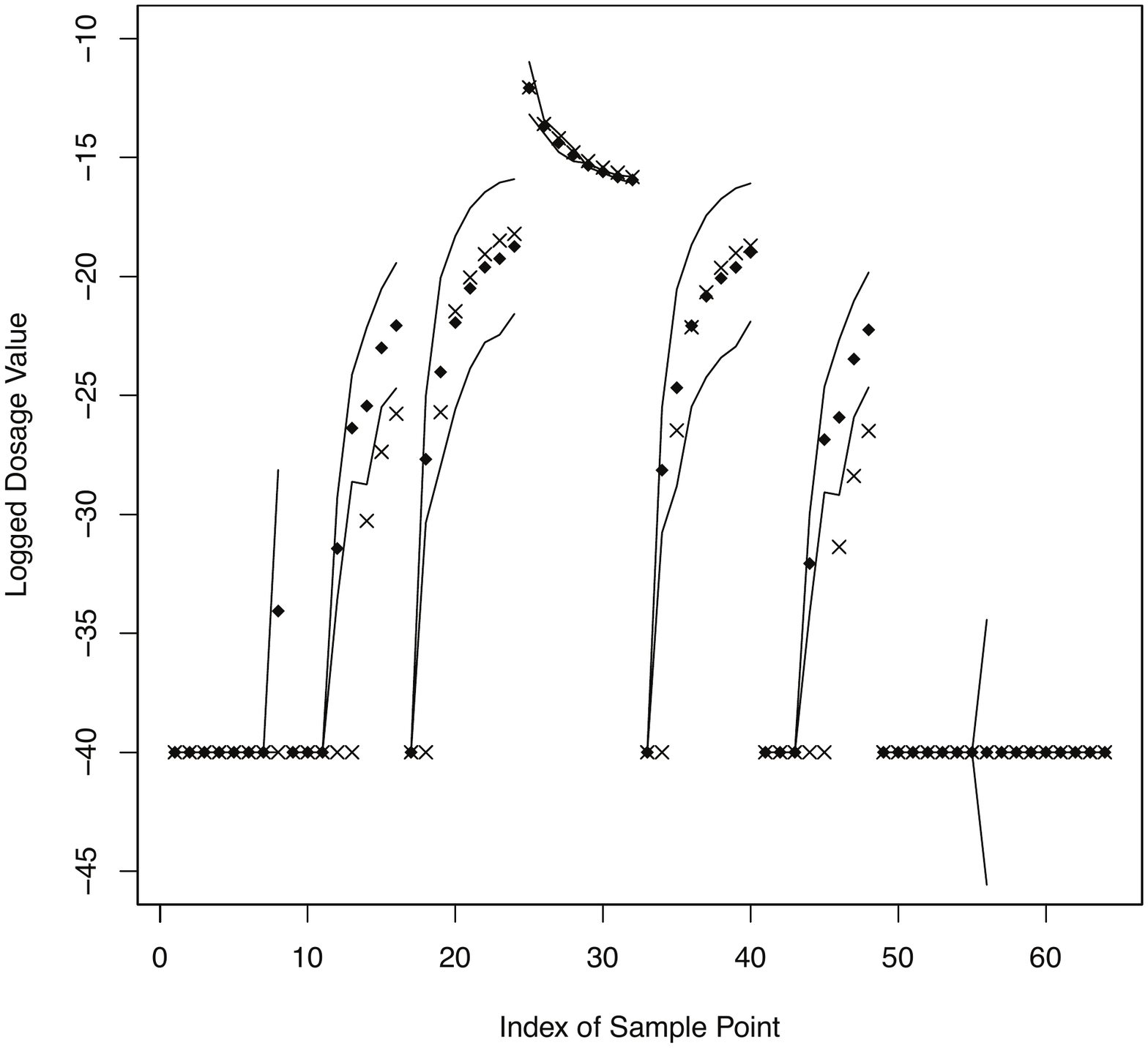} & \includegraphics[scale=\plotscale,clip=true]{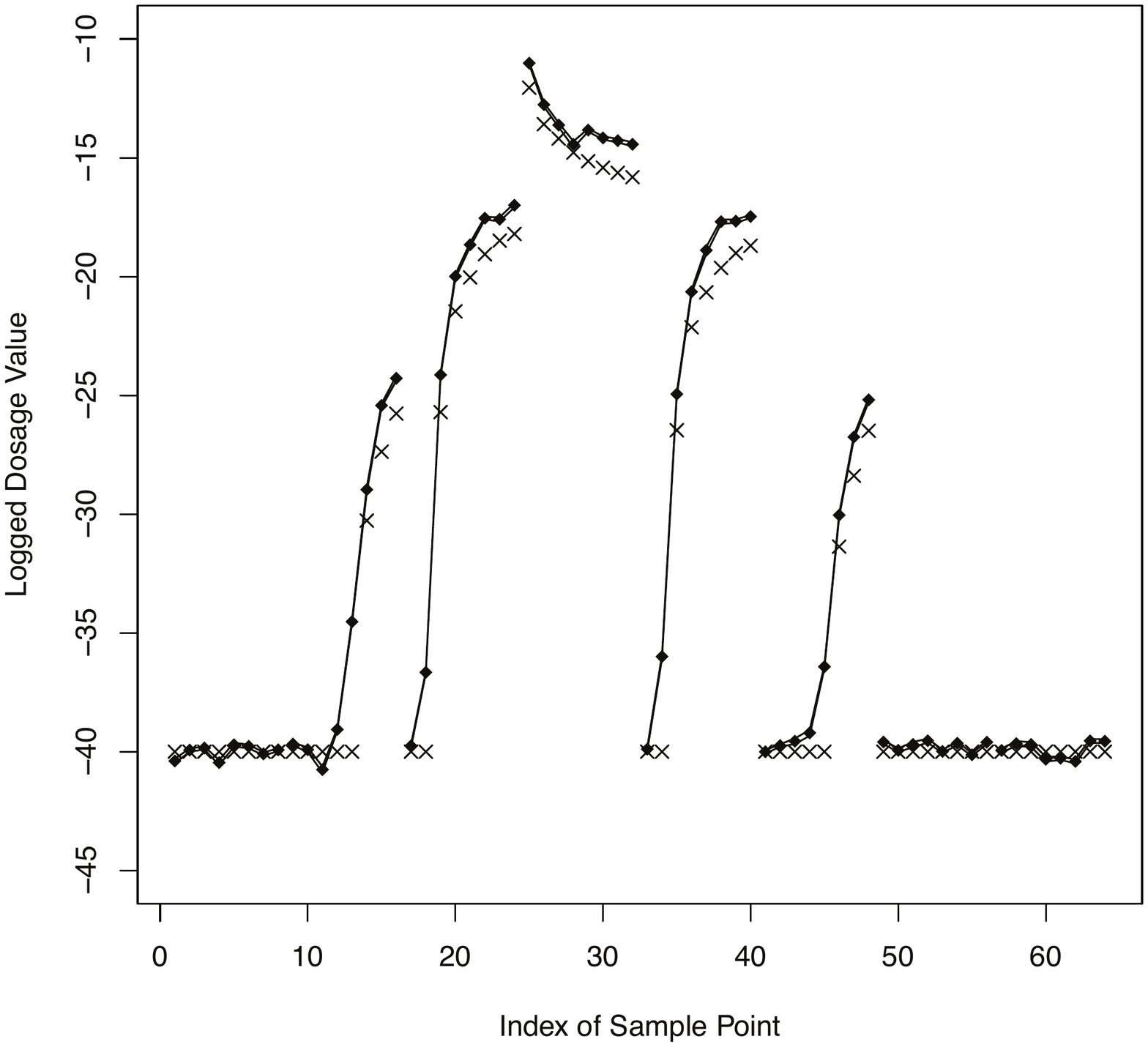} \\ [-8ex]
(c) & (d) \\[-14ex]
\includegraphics[scale=\plotscale,clip=true]{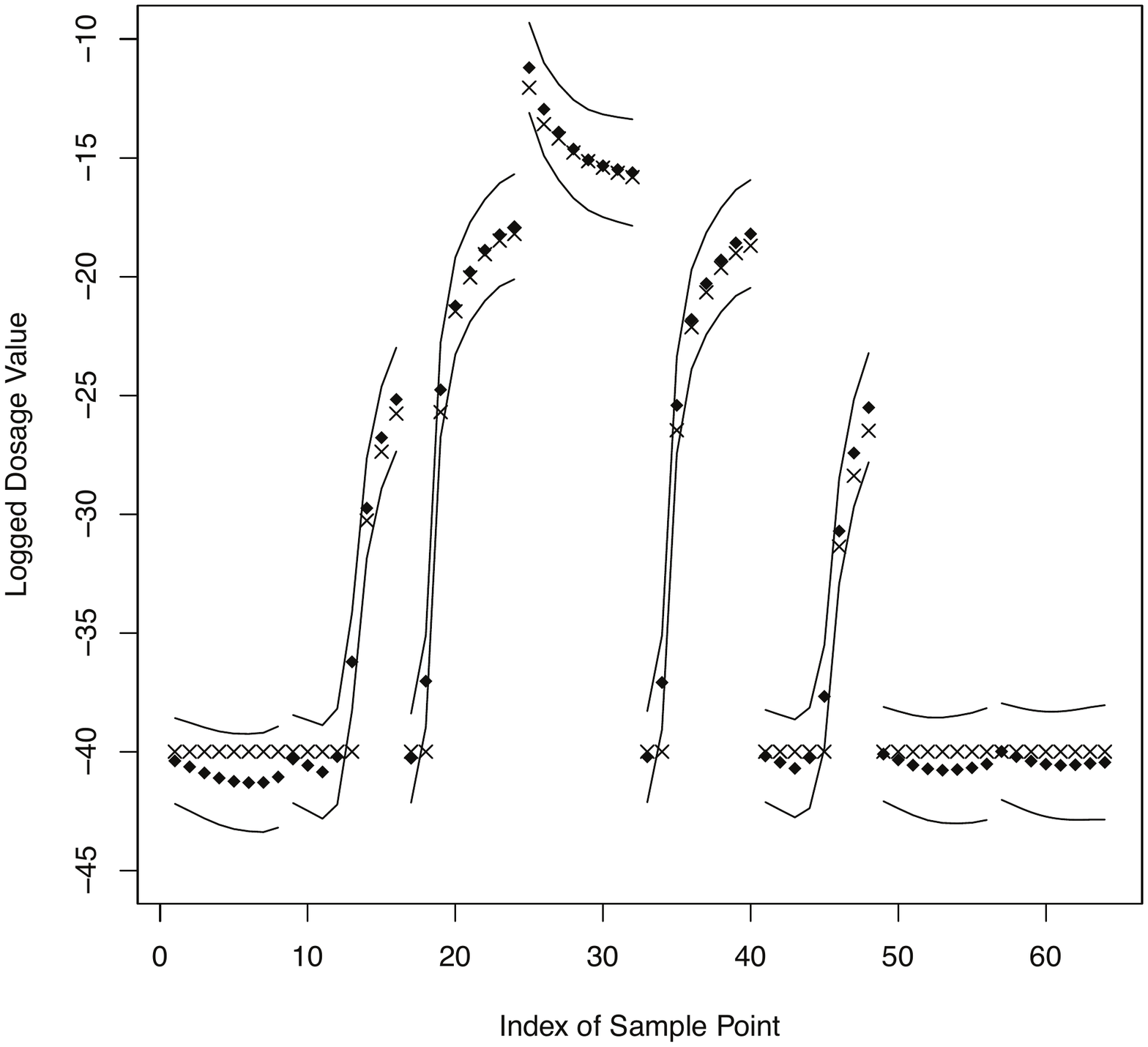} & \includegraphics[scale=\plotscale,clip=true]{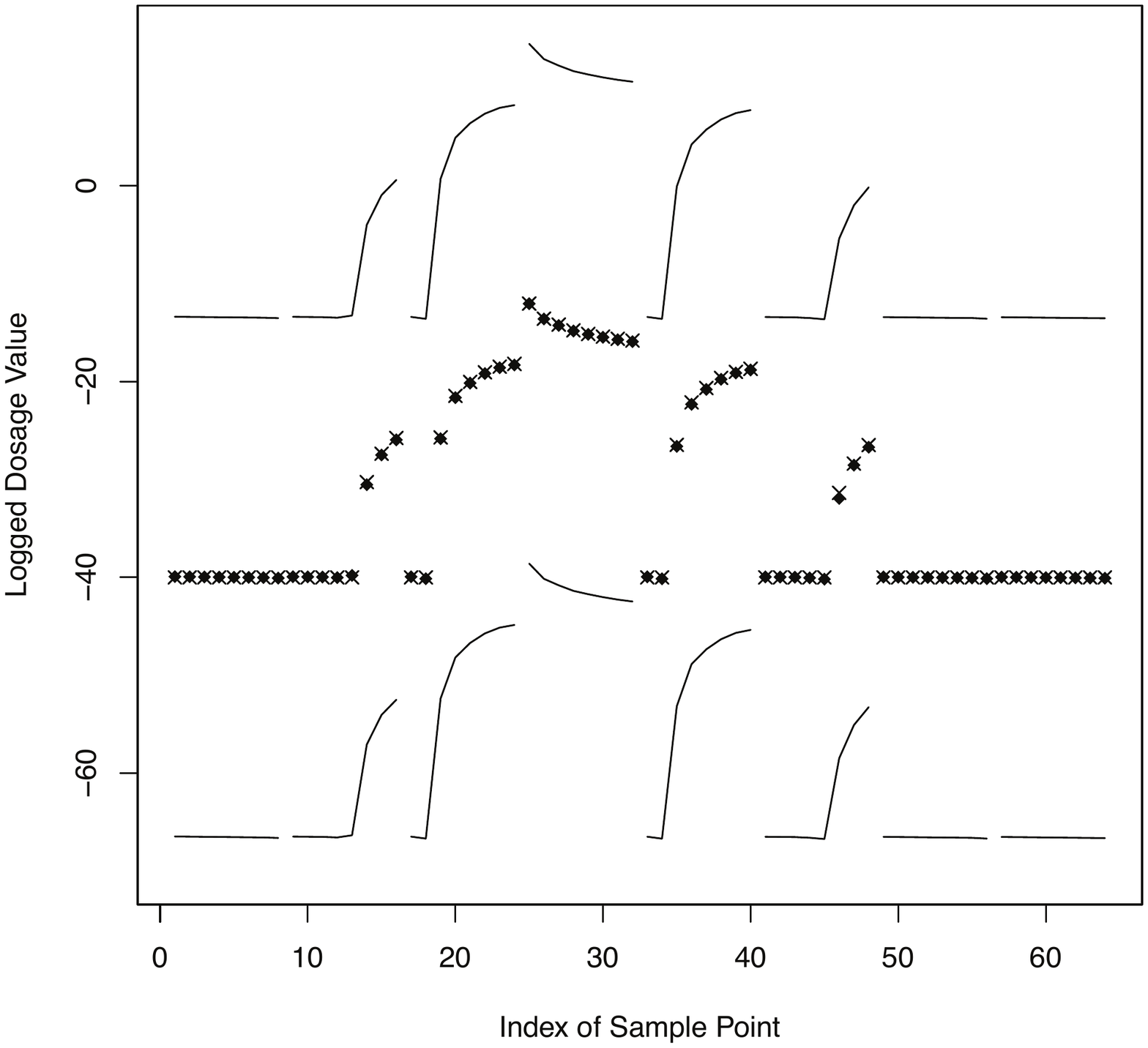} \\[-8ex]
\end{tabular}
\end{center}
\caption{\label{uncertainty} Logged dosage ($\times$), approximate posterior predictive mean ($\bullet$) and approximate 99\% predictive intervals for the $r=64$ spatial locations for one test run for the (a) PC-GP, (b) iTPRS, (c) sTPRS and (d) sGP emulators. The ordering of the output points corresponds to Figure~\ref{fancy}.}
\end{figure}

The Bayesian approach to emulation allows the uncertainty associated with predictions of the simulator to be quantified and assessed. Realistic and appropriate uncertainty quantification is a key determinant of a good emulator. Prediction uncertainty for the PC-GP and iTPRS-GP emulators can be obtained from the MCMC samples. For the sTPRS emulator, we use~(\ref{mvregpred}) with the conditional posterior distribution for $\bbeta$ and an additive error term corresponding to the approximation error~(\ref{errorprior}). A similar approach is adopted for the sGP emulator \cite{rougier2008}.

\begin{figure}[!t]
\begin{center}
\includegraphics[scale=.6,clip=true]{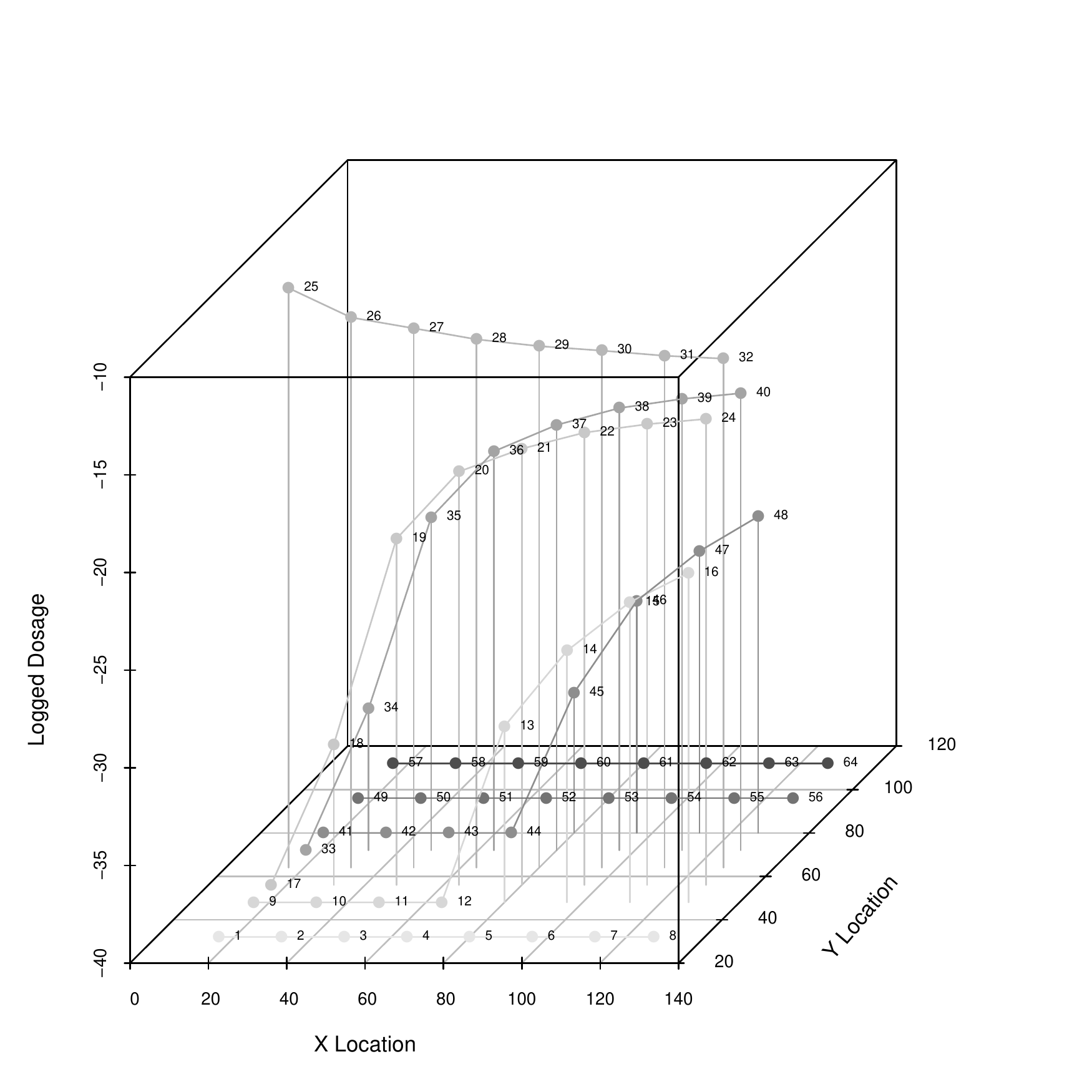}
\end{center}
\caption{\label{fancy} Relationship between the index of the sample output points and the spatial location.}
\end{figure}

To demonstrate the predicted uncertainty from the four emulators of the dispersion simulation, in Figure~\ref{uncertainty} we present prediction intervals (posterior predictive mean $\pm 3$ posterior predictive standard errors) for a single test run for each of the PC-GP, iTPRS-GP, sTPRS-GP and sGP emulators; other test runs produce similar results. The sample points in these figures are ordered by row, taking every 16th row and column from Figure~\ref{fig:persp}. They can be related to the output grid in Figure~\ref{fig:persp} by considering each set of 8 consecutive points as a horizontal transect across the grid. The locations of the points in the output space $\mathcal{S}$ are illustrated in Figure~\ref{fancy}. As would be expected in this application, the predictions have higher uncertainty in the more complex central areas of the output grid, with higher predictive means. 

We also calculate the frequentist coverage of these prediction intervals using the test data; we would expect coverage of around 99\%. From Figure~\ref{uncertainty},  it is clear that excluding within-run correlation in the iTPRS-GP emulator has led to substantial under-estimation of the uncertainty (coverage of 28\%), with almost all observed responses lying outside the, very narrow, prediction intervals. The MCMC chains had converged and sufficiently explored the parameter space, and hence the under-estimation of the uncertainty appears to be a consequence of modeling assumptions. The other emulators give more realistic assessment of uncertainty, with coverages of 83\% (PC-GP), 98\% (sTPRS-GP) and 100\% (sGP). The much wider predictive intervals for the sGP emulator plausibly result from the larger effective number of parameters in this emulator (with separate, dependent, Gaussian process models for each output point).

\subsection{Prediction on a high-resolution output grid}\label{dense}
In our exemplar simulation, the output grid covers a 10km by 10km spatial area. The low-resolution output grid consider up to this point can only accurately assess the performance of sensor locations at around a 1km resolution.  For application in a sensor placement tool, much higher resolution ($\sim$10m) is required. To assess the performance of the sTPRS-GP and sGP emulators at a high resolution, we define a $63\times 63$ regular lattice as an output grid (with no boundary points) and attempt to predict the output for untried $\bx$ on this much finer scale. We choose these two emulators because of their superior performance on the low-resolution output grid.  

\begin{figure}[!t]
\begin{center}
\includegraphics[scale=.4,clip=true]{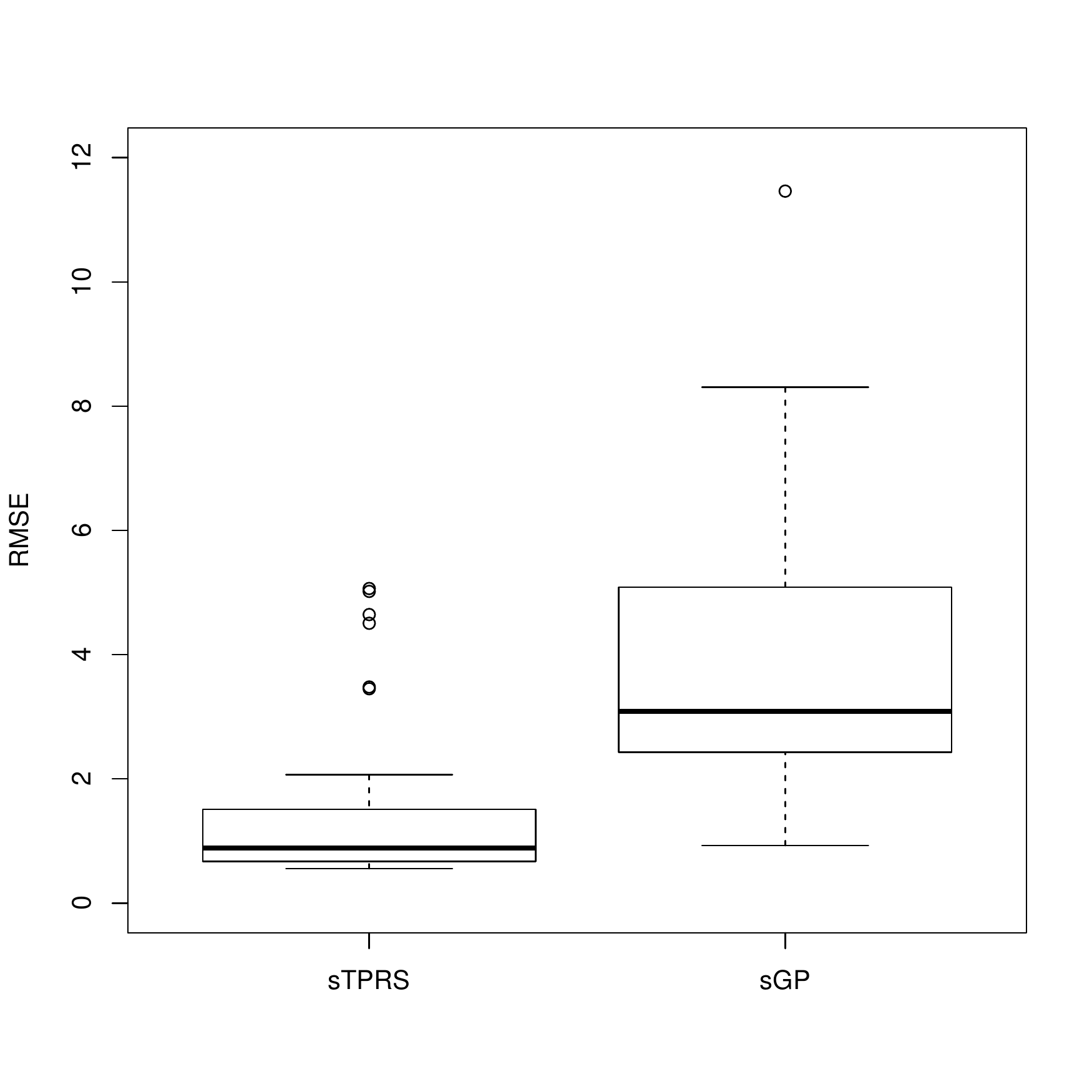}
\end{center}
\caption{\label{fig:MSEBig}{RMSE for the sTPRS-GP and sGP emulators for the high-resolution output grid using the posterior predictive mean across the test set.}}
\end{figure}

For this size output grid ($r=3639$), it is no longer computationally feasible to train the sGP emulator using the whole output grid in realistic time (it requires multiple inversions of a $3696\times3696$ correlation matrix). However, as the sGP is a model for functional data, instead we train the emulator using an $8\times 8$ sub-grid (actually the same output grid as in the previous sections) and then predict at the interim locations. The sTPRS-GP emulator, however, performs dimension reduction on the output grid and hence can be trained using much larger data sets, with the basis functions defined using the complete output grid. The benefit of this method is two-fold: prediction is actually computationally less expensive and all of the output data can be used, providing more accurate prediction across the whole output domain $\mathcal{S}$.  The RMSE of the two approaches are shown in Figure~\ref{fig:MSEBig}.  

\begin{figure}[!t]
\centering
\subfigure[Contour Plot]{
\includegraphics[scale=0.32]{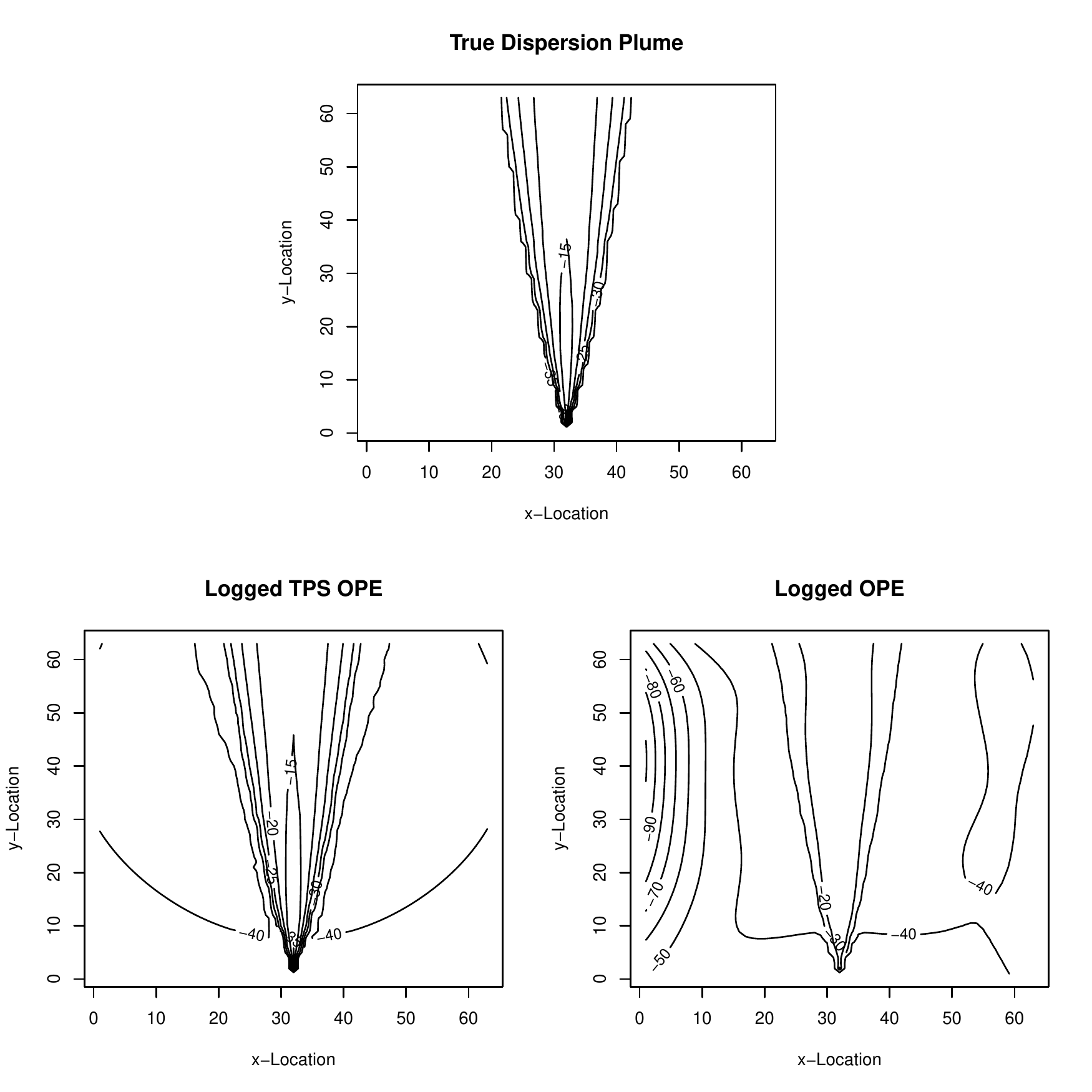}
}
\subfigure[Index Plot]{
\includegraphics[scale=0.32]{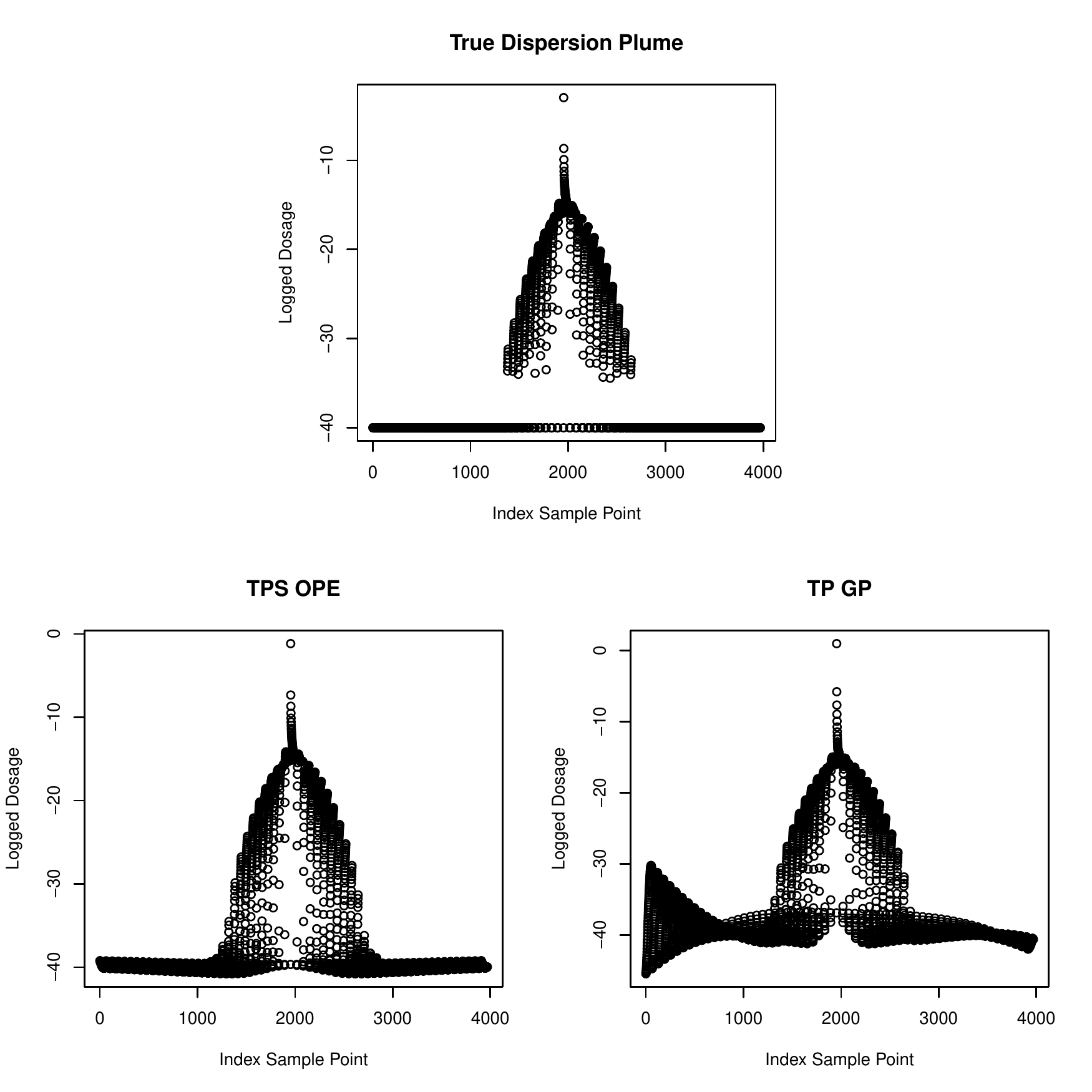}
\label{fig:actcontour2}
}
\label{fig:actualPlume2}
\caption{\label{fig:figBoxBig} Contour plot and index logged dosage to compare the sTPRS-GP and sGP emulators.}
\end{figure}

It is clear from this figure that the sTPRS emulator provides substantially more accurate predictions that the sGP emulator. An important advantage here of the sTPRS-GP emulator is that it does not need to use an estimated correlation structure on the output grid for prediction, unlike the sGP emulator.  When the output grid displays non-stationary correlation, as with this dispersion example, the sGP emulator struggles to accurately predict at interim locations, see Figure ~\ref{fig:figBoxBig} for an illustration. In general, for detailed output domains, e.g. urban locations, using a low-resolution output grid to train the sGP emulator could result in missing important details in the response.

\section{Artificial example}\label{artex}

The dispersion example demonstrated the effectiveness of the sTPRS-GP emulator for prediction on dense grids with changing spatial variation. In this section, we use an artificial environmental example to further compare the sTPRS-GP and sGP emulators.

The example is a modified version of a simulator that has previously been used to demonstrate calibration methodology \cite{brsrwm2008}. The simulator models a chemical pollutant spill at two locations into a long, narrow holding channel. We assume the location and time of both spills is known and fixed but that the mass and diffusion rate at each spill location may be inputs to the simulator. For the pollutant concentration at location $s_1$ and time $s_2$ (so $q = 2$), the simulator has the simple closed form
\begin{multline}\label{artsim}
Y(\bs, \bx) = \frac{x_1}{\sqrt{4\pi x_2 s_2}}\exp\left(\frac{-s_1^2}{4x_2 s_2}\right) \\ 
+ \frac{x_3}{\sqrt{4\pi x_4 (s_2 - \tau)}}\exp\left(\frac{-(s_1-L)^2}{4x_4 (s_2 - \tau)}\right)\mathbb{I}(s_2 > \tau)\,,
\end{multline}
where $x_1, x_2$ and $x_3, x_4$ are the mass and diffusion rate of spilled pollutant at spills 1 and 2, respectively. The first spill is at (location, time) = $(0,0)$; the second is at $(L, \tau) = (1.505, 30.1525)$.

\newcommand{\environplotscale}{0.33} 
\begin{figure}[!t]
\begin{center}
\begin{tabular}{cc}
(a) & (b) \\[-5ex]
\includegraphics[scale=\plotscale,clip=true]{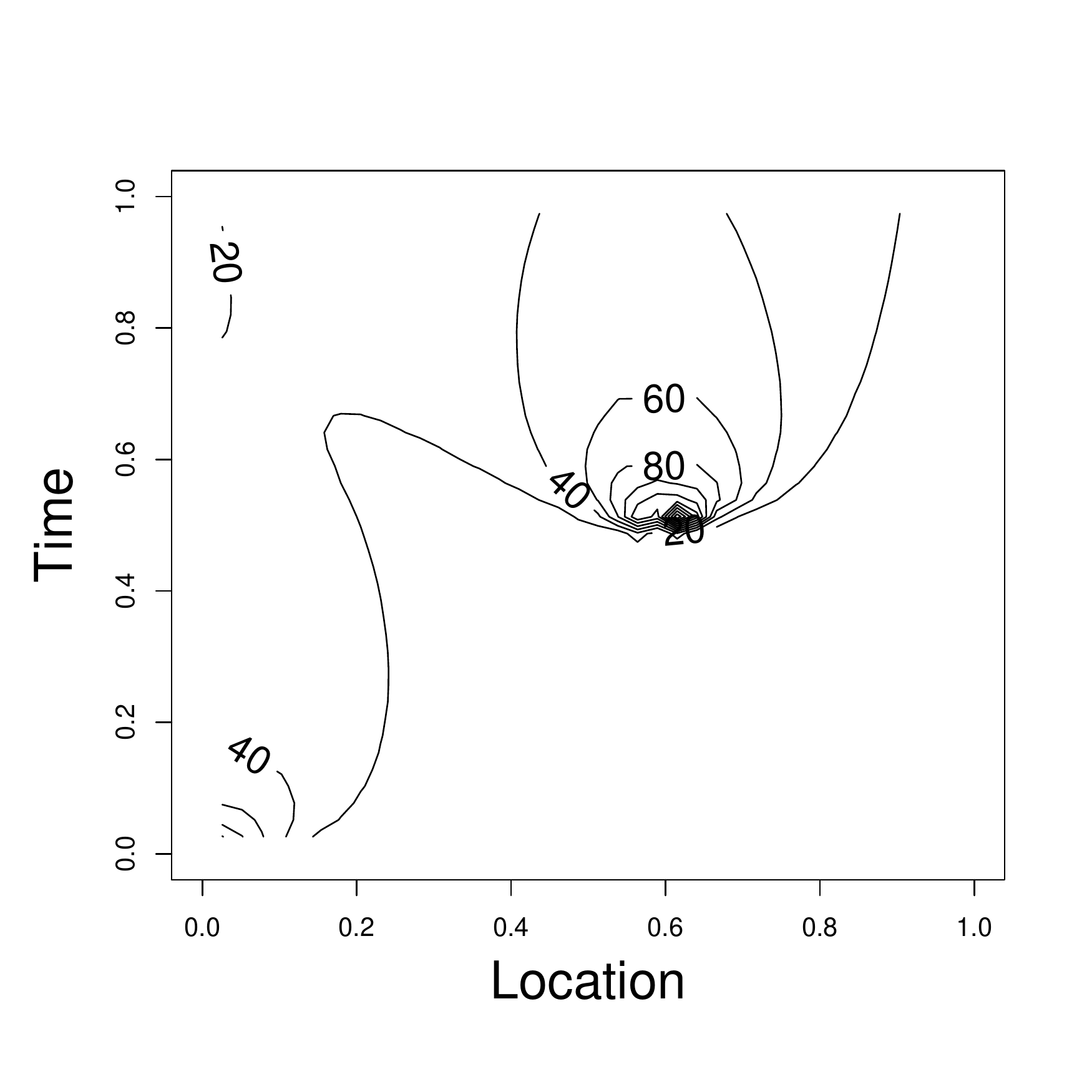} & \includegraphics[scale=\plotscale,clip=true]{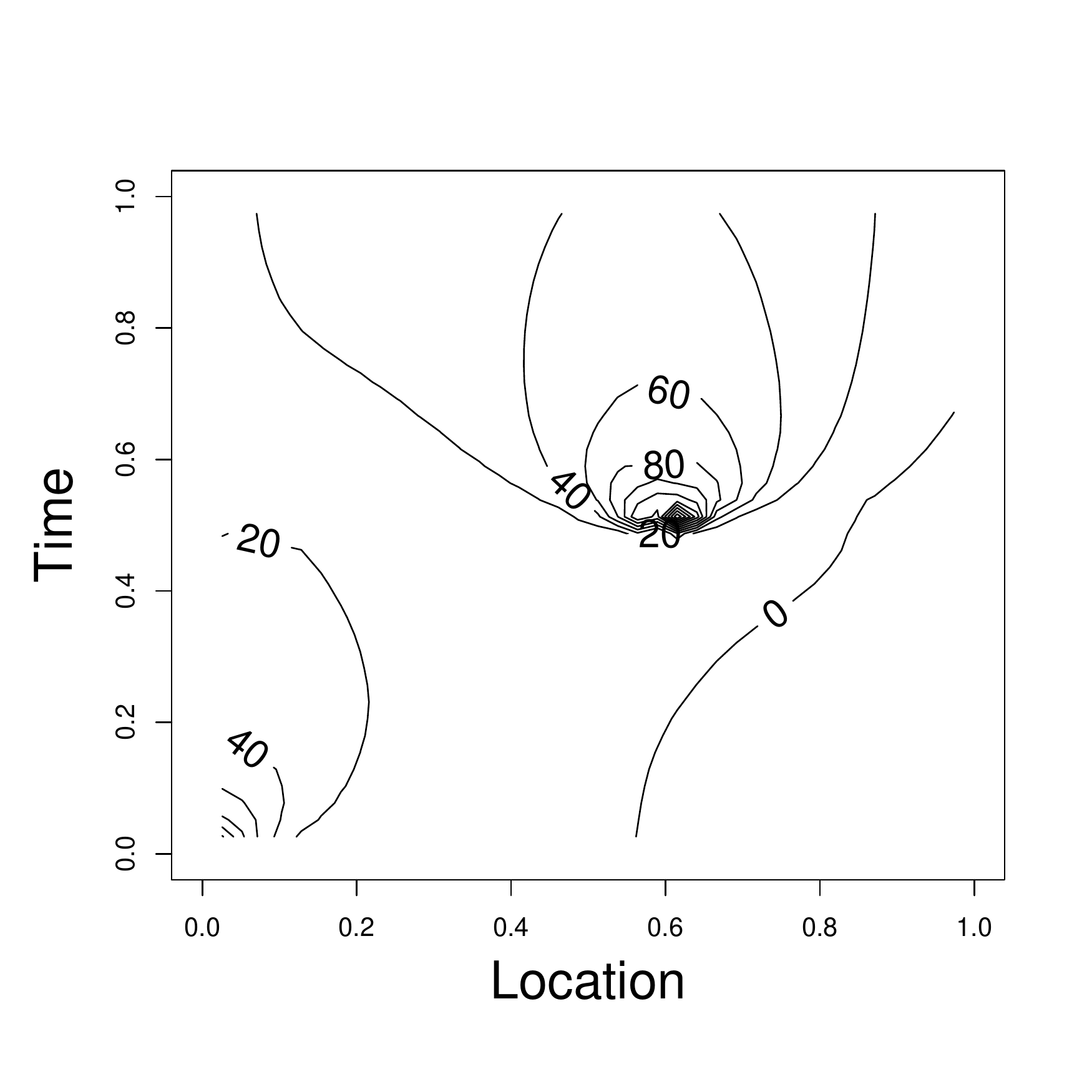} \\[-1ex]
\multicolumn{2}{c}{(c)} \\[-5ex]
\multicolumn{2}{c}{\includegraphics[scale=\plotscale,clip=true]{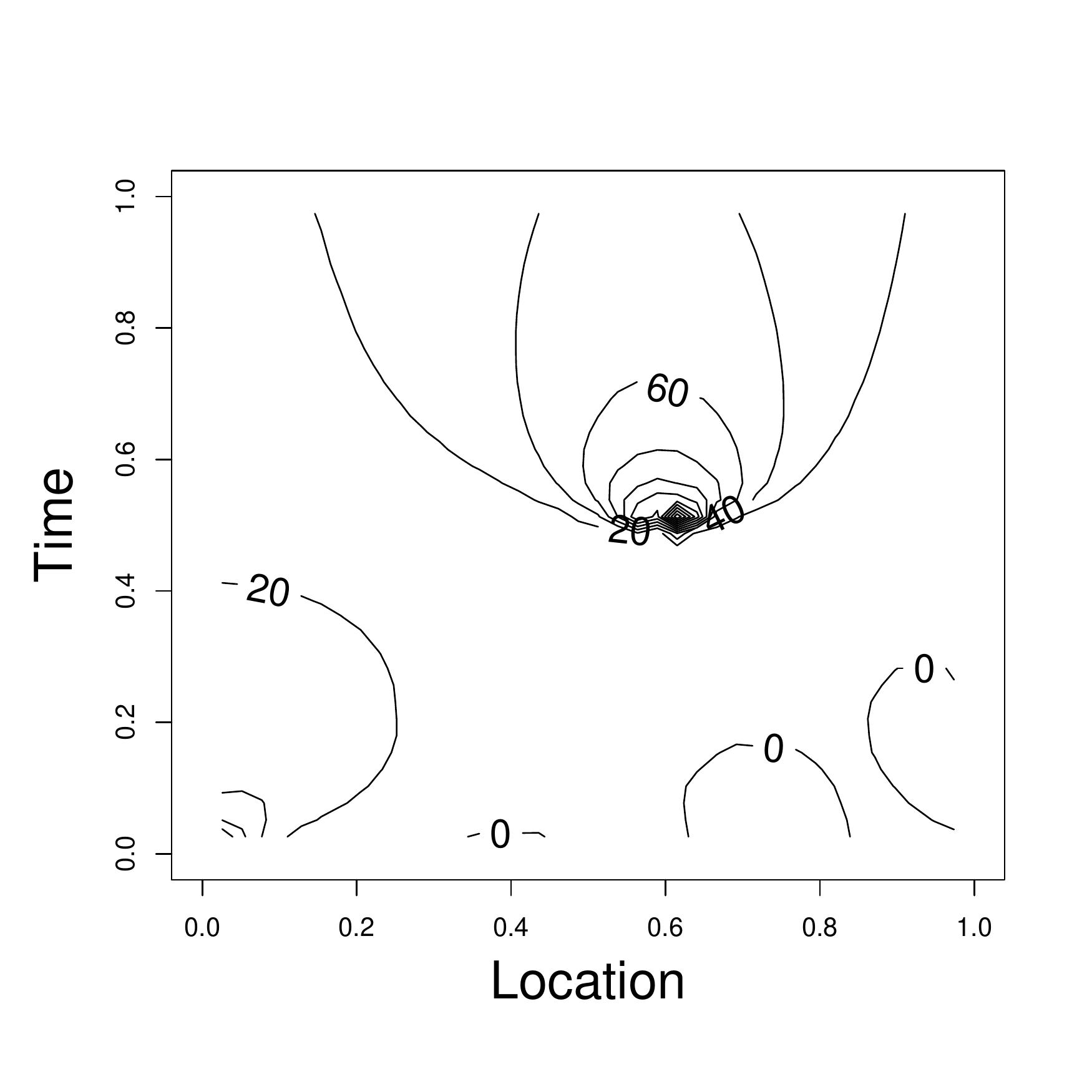}} \\[-5ex]
\end{tabular}
\end{center}
\caption{\label{environ1} Contour plots for the artificial environment example showing (a) the true concentration surface from the simulator and the predicted concentration surface from (b) the sTPRS-GP emulator and (c) the sGP emulator.}
\end{figure}

We generated data from simulator~\eqref{artsim} for four scenarios, corresponding to $d = 1, \ldots, 4$ variables, starting with varying just the first mass $x_1$, then the first mass and diffusion rate $(x_1, x_2)$, and so on. Variables held fixed were set to the mid-points of their ranges, $7\le x_1, x_3 \le 13$ and $0.02\le x_2, x_4 \le 0.12$. We adapted code from \texttt{http://www.sfu.ca/$\sim$ssurjano} and used four $n=80$ run maximin Latin Hypercube designs with $q =1, \ldots, 4$ to generate the training data, and separate samples from continuous distributions to generate validation and test data sets, each with 10 runs. 

We took the most effective emulators from the dispersion example, sTPRS-GP and sGP, and applied them to the four scenarios. For each scenario, the validation data was used to choose the various prior hyper-parameters for each emulator, including the number, $p$, of functions in the thin-plate regression spline basis. The sTPRS-GP emulator was estimated using an output grid with $r=2500$ location/time points, whereas the sGP emulator was estimated using a much smaller grid with $r=100$ points. Fitting the sGP emulator to the larger output grid was computationally infeasible. In each case, a $r=2500$ output grid was used for the validation and test sets. All output grids were regular two-dimensional lattices with no boundary points. We again scaled the training, validation and test data sets to have concentration with mean zero and standard deviation 1 at each output point. 

Figure~\ref{environ1} shows some typical output from this modeling exercise when $d = 4$. In this case, both the sTPRS-GP and sGP emulators capture the basic features of the true concentration surface. However, the sGP emulator, which was estimated using the much smaller output grid, has overestimated the mass and/or diffusion at both pollutant spills, resulting in higher predicted concentrations than the sTPRS-GP, which better captures the true concentration. Here, the sTPRS-GP has average root mean squared error 7\% smaller than the TP-GP.

\begin{figure}[!t]
\begin{center}
\begin{tabular}{cc}
(a) & (b) \\[-5ex]
\includegraphics[scale=\plotscale,clip=true]{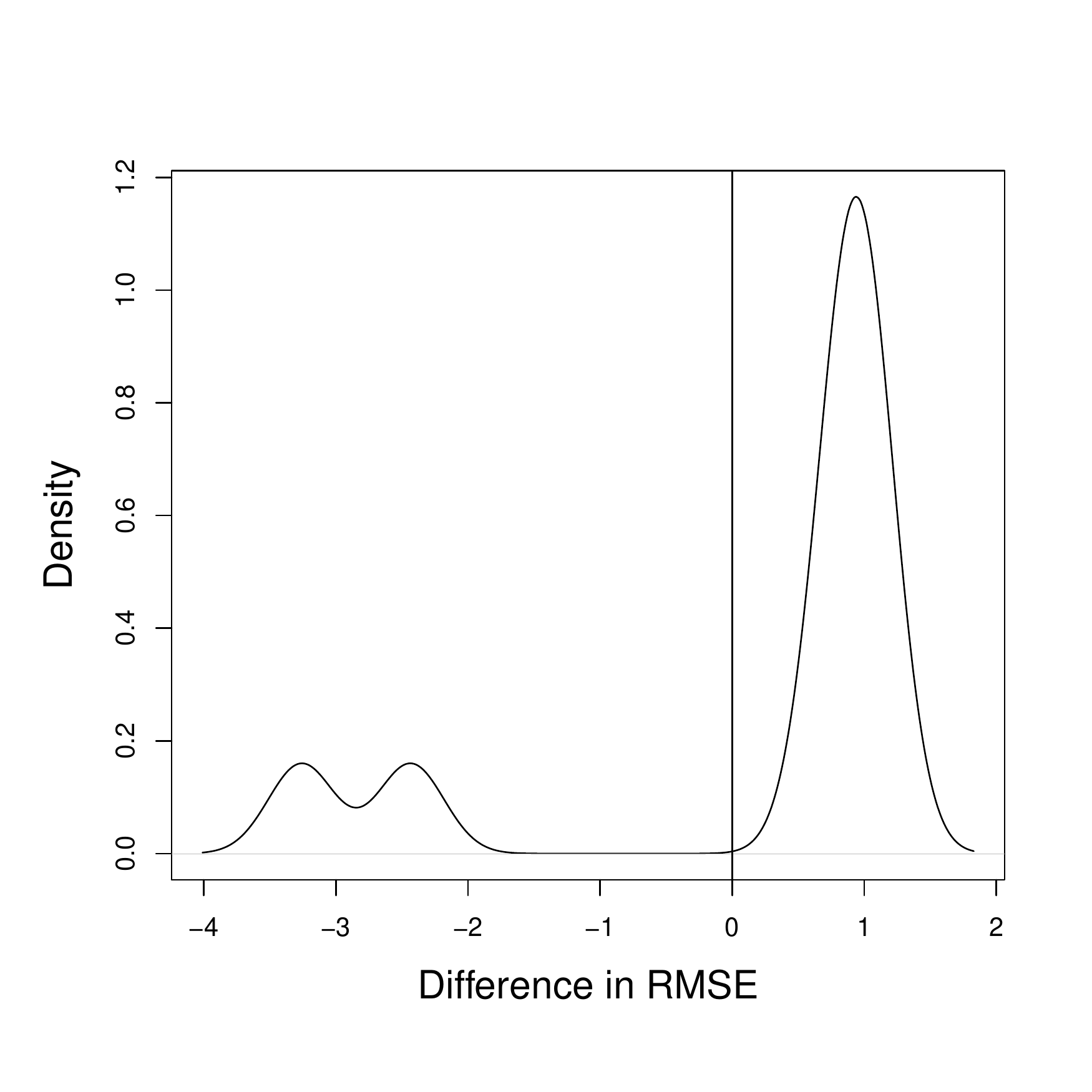} & \includegraphics[scale=\plotscale,clip=true]{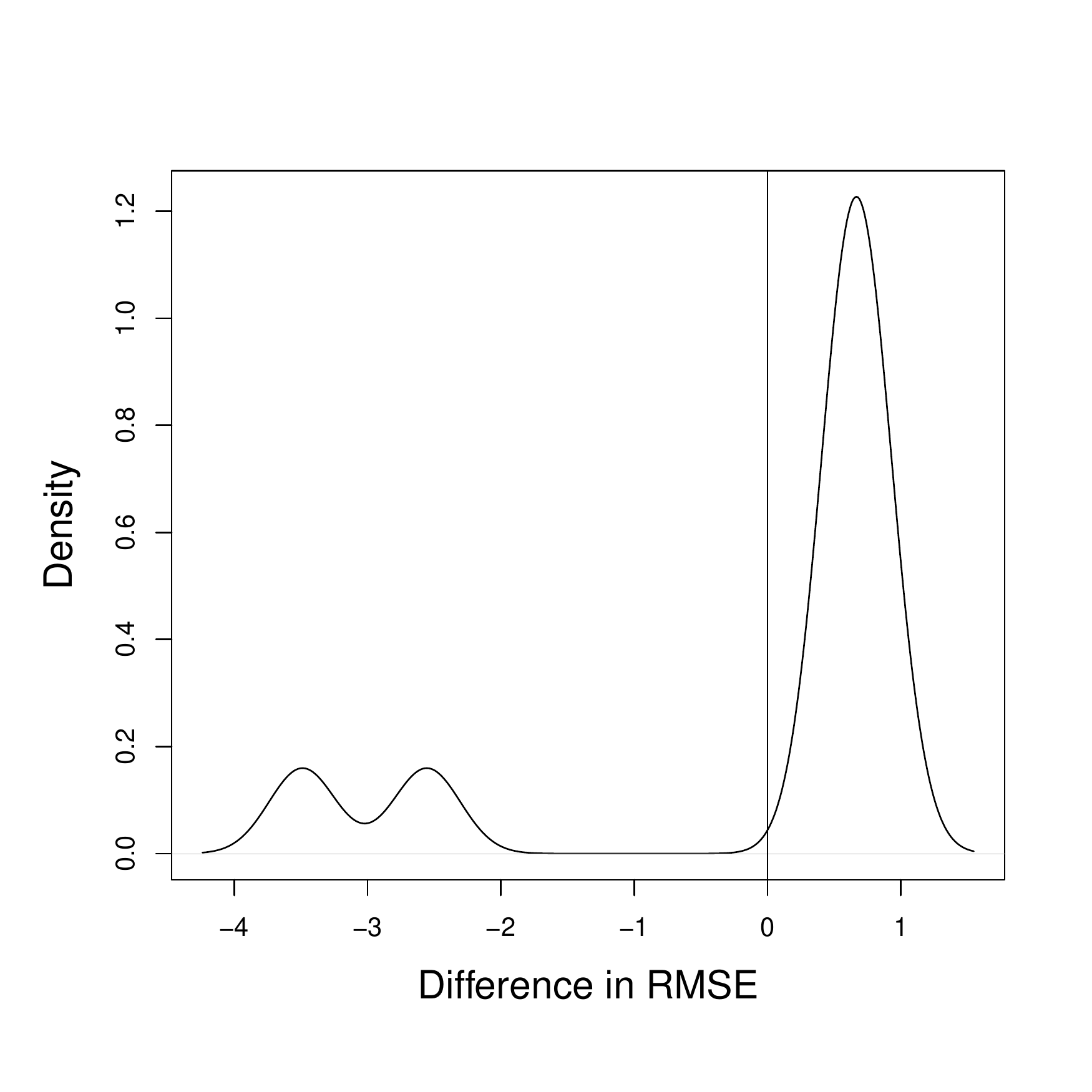} \\[-1ex]
(c) & (d)\\[-5ex]
\includegraphics[scale=\plotscale,clip=true]{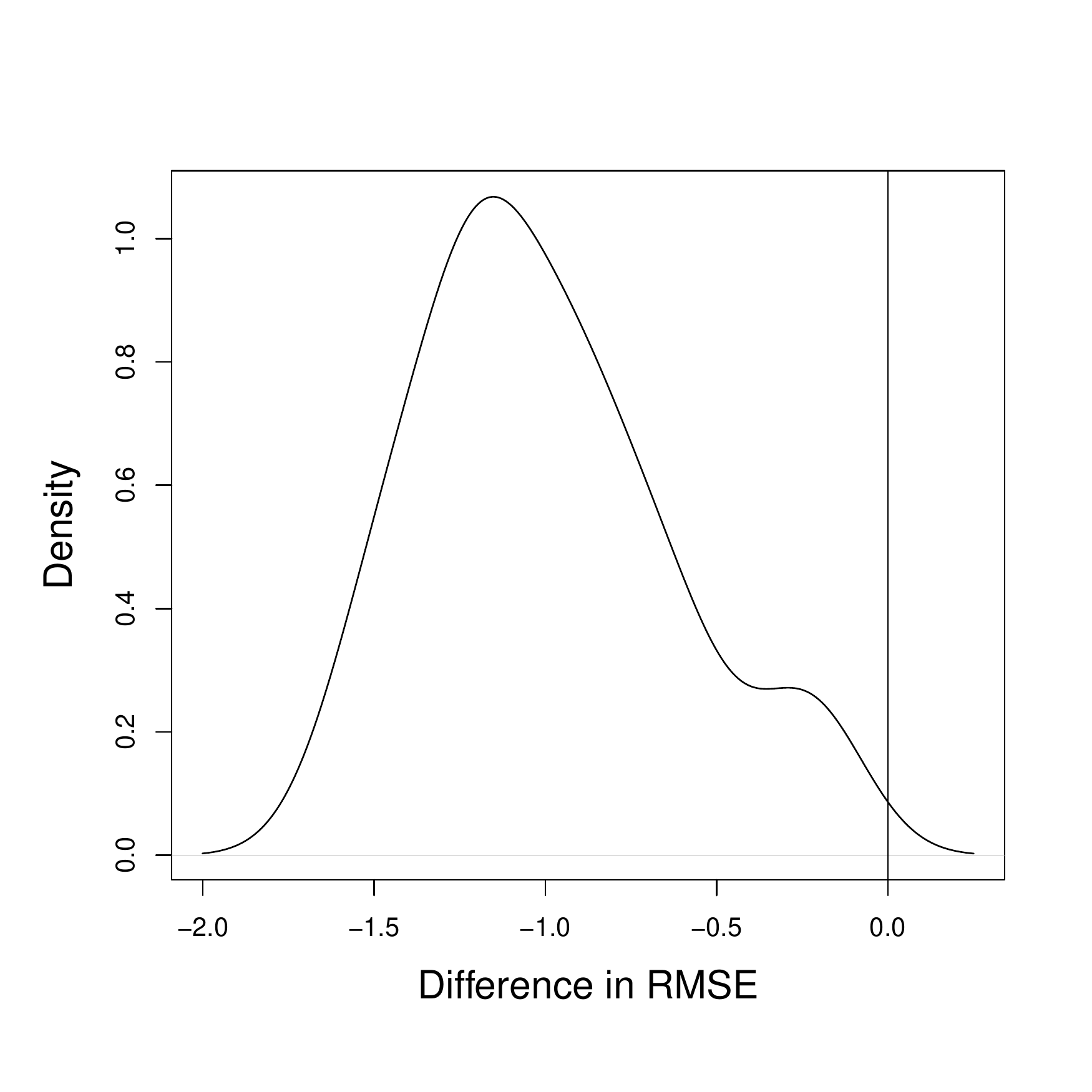} & \includegraphics[scale=\plotscale,clip=true]{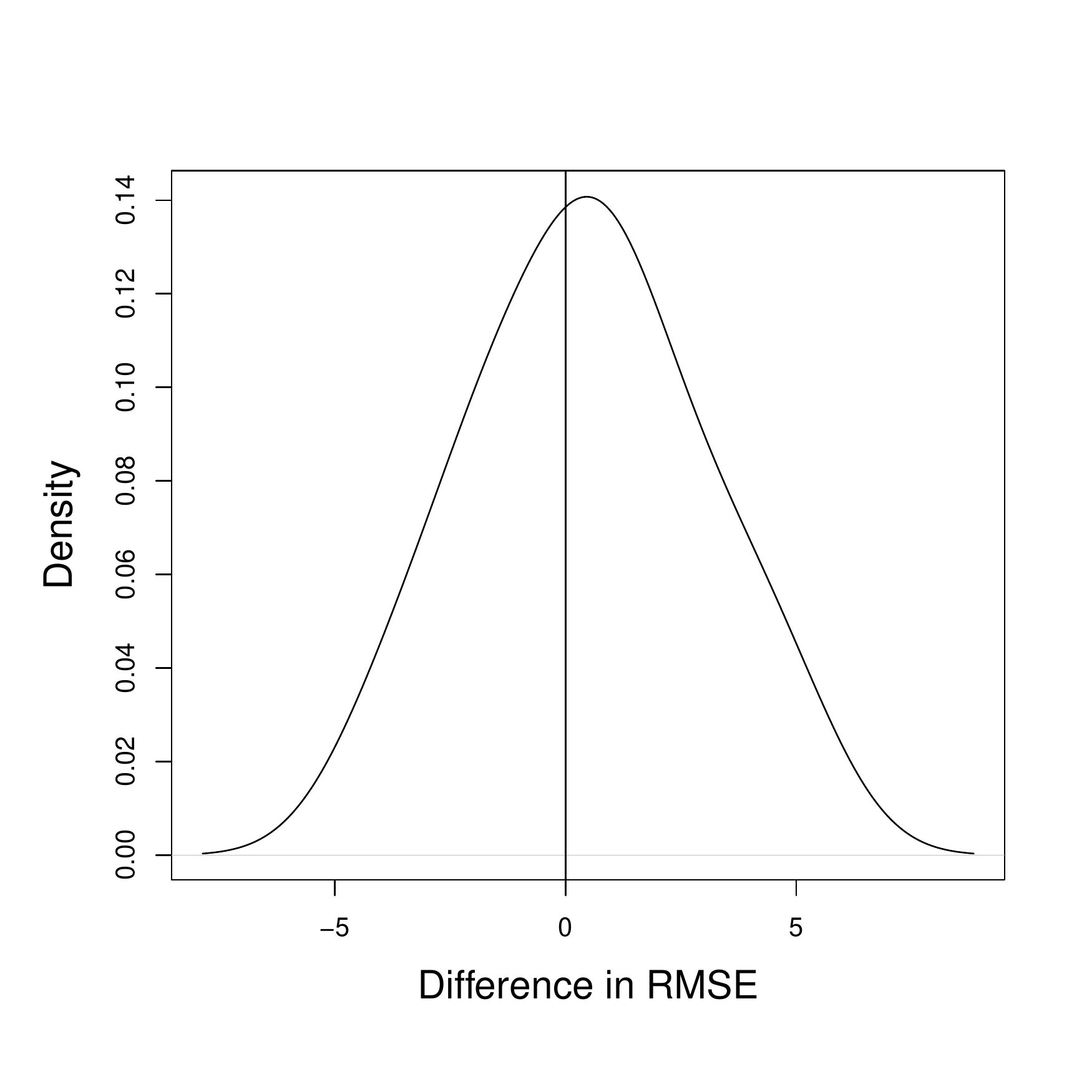} \\[-5ex]
\end{tabular}
\end{center}
\caption{\label{environ2} Density plots of the difference in test set RMSE between the sGP and sTPRS-GP emulators for the environmental simulator with (a) one, (b) two, (c) three and (d) four variables.}
\end{figure}

For fewer input variables ($d<4$), the sGP emulator had a predictive advantage in terms of RMSE, see Figure~\ref{environ2}. For $d = 1, 2, 3$, the improvement in average RMSE for the sGP was between 2\% and 26\%, with the sTPRS-GP performing comparatively poorest for $d = 3$. While there seems to be little pattern in performance here, it is plausible that the improved performance of the TPRS-GP emulator for $d=4$ is due to (i) greater variation in the response across the output grid, particularly when there are high levels of concentrated mass at the two sources (high mass, low diffusion rate) and (ii) more detailed changes in the response surface occurring between runs. Both these situations occur more often when both diffusion rates are varied. 

\section{Discussion and future work}
\label{disc}

We have presented the emulation of multivariate simulators with two-dimensional output structure. Thin-plate regression splines were demonstrated to be an effective method for dimensional-reduction, that resulted in substantially increased prediction accuracy over the use of principal components. When a separable covariance structure is assumed for the basis coefficients, a computationally feasible emulator results that provides realistic measures of uncertainty. For the separable emulator, we have adopted a plug-in approach using the posterior predictive distribution conditional on the correlation parameters. Clearly, there is the potential for this approach to under-estimate the posterior uncertainty. However, the results in Section~\ref{uncertaintysec} suggest that it is more important to account for the within-run correlations resulting from the non-orthogonal thin-plate spline basis functions.

When high-resolution prediction was required in the dispersion example, the thin plate regression methodology proved more accurate and computationally feasible than functional data modelling using a separable Gaussian process model. A key reason for the advantage of the TPRS approach is the non-stationary nature of the output data, with differing correlation lengths across the output domain. The artificial example demonstrated that the two methodologies can perform very similarly, although sTPRS-GP emulator was again preferred when non-stationary output variation was observed. An area for future research is to define more fully the scenarios (non-stationary correlation, high-resolution spatial features) in which the TPRS approach has a predictive advantage. Other research could include developing multivariate emulators using Gaussian processes with non-stationary or non-separable covariance structures \cite{FOU2013},  modeling dispersion simulators with qualitative inputs (e.g. \cite{qian2008}) such as release type relating to the source term, and building emulators for dynamic responses such as concentration over time. Further research is also required into designing the computer experiments for these multivariate hierarchical problems, including the choice of the simulator inputs and also, in some applications, the selection of output domain locations (cf \cite{bdw}).\\[3ex]

\section*{Acknowledgments}

This work was supported by the Defense Threat Reduction Agency (grant HDTRA1-08-1-0048), a Dstl Associate Fellowship for V.E. Bowman and an EPSRC Fellowship (EP/J018317/1) for D.C. Woods.

Content includes material subject to \textcopyright~Crown copyright (2015), Dstl. This material is licensed under the terms of the Open Government Licence except where otherwise stated. To view this licence, visit \texttt{http://www.nationalarchives.gov.uk/doc/open-government-licence/version/3} or write to the 
\newline Information Policy Team, 
\newline The National Archives, 
\newline Kew, 
\newline London TW9 4DU, 
\newline or email: \texttt{psi@nationalarchives.gsi.gov.uk}.

\appendix
%
%
%
%
%
%
%
%
%
%
%
%
%
%
%

\section{Development of the conditional posterior density for the sTPRS-GP emulator}
\label{seppost}

For $\bbeta | \tau, C \sim N(\boldsymbol{0}_{np},C\tau)$, a mean-zero Gaussian process,  



$$
\bbeta^\star | \bbeta, \tau \sim N(\boldsymbol{m}, \tau S)\,,
$$

\noindent with $\boldsymbol{m} = \left(C^\star\right)^\mathrm{T}C^{-1}\bbeta$, and $S=C^{\star\star} - \left(C^\star\right)^\mathrm{T}C^{-1}C^\star$. Here, $C^\star$ is the $np\times p$ matrix of correlations between $\bbeta$ and $\bbeta^\star$, and $C^{\star\star}$ is the $p\times p$ matrix of correlations between the elements of $\bbeta^\star$. 

Hence, for a separable correlation structure with with $C = W\otimes V$, point estimates of $\bbeta^\star$ can be provided by

\begin{eqnarray*}
E(\bbeta^\star | \bbeta, \tau) & = & \left(\boldsymbol{w}^{\star}\otimes V\right)^\mathrm{T}W\otimes V\bbeta\,, 
\end{eqnarray*}

\noindent with $\boldsymbol{w}^{\star}$ the $n$-vector of correlations between $\bx^\star$ and $\bx$. Similarly, uncertainty in $\bbeta^\star$ can be assessed via

\begin{equation}\label{betastarvar}
\mbox{Var}(\bbeta^\star | \bbeta,\tau) = \tau\left\{C^{\star\star}- V\otimes\left(\boldsymbol{w}^{\star}\right)^\mathrm{T}\left(W^{-1}\otimes V^{-1}\right)\boldsymbol{w}^{\star}\otimes V\right\}\,. 
\end{equation}

In addition, 

$$
\pi(\tau | \bbeta)\propto \tau^{-1-(a_\tau+np)/2}\exp\left\{-(b_\tau+\bbeta^\mathrm{T}C^{-1}\bbeta)/2\tau^{-1}\right\}\,.
$$ 

\noindent That is, $\tau^{-1} | \bbeta\sim \mbox{Gamma}(a_\tau+np,(b_\tau+\bbeta^\mathrm{T}C^{-1}\bbeta)^{-1})$.  A plug-in estimate of the posterior uncertainty in $\bbeta$ can be achieved by replacing $\tau$ in~(\ref{betastarvar}) with its posterior mean, $E(\tau | \bbeta) = \left(b_\tau+\bbeta^\mathrm{T}W^{-1}\otimes V^{-1}\bbeta\right)/(a_\tau+np-2)$.

\section{Construction of the sampling distribution for the iTPRS-GP emulator}\label{itpsappendix}

From equation~(\ref{betaprior}),

\begin{equation}\label{appbeta}
\bbeta | C \sim N(\boldsymbol{0}_{np},C)\,, 
\end{equation}

\noindent where $C$ is an $np\times np$ covariance matrix determined by the specified covariance structure and parameter vectors $\btau$ and $\btheta$. We specify independent $\mbox{Gamma}(a_{\tau},b_{\tau})$ priors for each $\tau_{i}$ and independent Beta$(a_{\theta},b_{\theta})$ priors for each $\theta_{i}$

\begin{eqnarray*}
\pi(\tau_{i}) &\propto& \tau_{i}^{a_{\tau}-1}\exp ^{-\tau_{i}/b_{\tau}} \quad\quad i = 1,\ldots, pn\,, \\
\pi(\theta_{ik}) &\propto& \theta_{ik}^{a_{\theta}-1}(1-\theta_{ik}) ^{b_{\theta}-1} \quad\quad i = 1,\ldots, pn;\, k=1\ldots dp\,.
\end{eqnarray*}


We define an $nr$ vector $\tilde{\bY}$ to be the concatenation of all $n$ simulation output vectors

\begin{equation*}
\tilde{\bY} = \mbox{vec}([\bY_{1};\ldots;\bY_{n}])\,.  
\end{equation*} 

\noindent Given the precision $\sigma^{-2}$ of the errors the likelihood is then

\begin{equation*}
L(\tilde{Y}|\bbeta,\sigma^2) \propto \sigma^{-nr}\exp\{-\frac{1}{2}\sigma^{-2}(\tilde{\bY}-A\boldsymbol{\beta})^{T}(\tilde{\bY}-A\boldsymbol{\beta})\}\,,
\end{equation*}

\noindent where the $nr\times np$ matrix 

\begin{equation*}\label{Aeq}
A=[I_{n}\otimes \ba_{1};\ldots;I_{n}\otimes \ba_{p}]
\end{equation*}

\noindent is constructed from the basis vectors $\ba_k$.  A $\mbox{Gamma}(a_{\sigma},b_{\sigma})$ distribution is specified for the error precision $\sigma^{-2}$. 

We then follow the factorization in \cite{hgwr},

\begin{eqnarray*}
L(\tilde{Y}|\bbeta,\sigma^2) &\propto& \sigma^{-np}\exp\{-\frac{1}{2}\sigma^{-2}(\bbeta-\hat{\bbeta})^{T}(A^{T}A)(\bbeta-\hat{\bbeta})\} \times\\\nonumber
&&\quad\sigma^{-n(r-p)}\exp\{-\frac{1}{2}\sigma^{-2}\tilde{\bY}^{T}(I-A(A^{T}A)^{-1}A^{T})\tilde{\bY}\}\,,
\end{eqnarray*}

\noindent to define a dimension reduced likelihood and a modified $\mbox{Gamma}(a'_{\sigma},b'_{\sigma})$ prior for $\sigma^{-2}$:

\begin{eqnarray*}
L(\hat{\bbeta}|\bbeta,\sigma^2) &\propto& \sigma^{-np}\exp\{-\frac{1}{2}\sigma^{-2}(\hat{\bbeta}-\bbeta)^{T}(A^{T}A)(\hat{\bbeta}-\bbeta)\}\,,\\\nonumber
a'_{\sigma}&=& a_{\sigma}+\frac{n(r-p)}{2}\,,\\\nonumber
b'_{\sigma}&=& b_{\sigma}+\frac{1}{2}\tilde{\bY}^{T}(I-A(A^{T}A)^{-1}A^{T})\tilde{\bY}\,,\\\nonumber
\hat{\bbeta}& =& (A^{T}A)^{-1}A^{T}\tilde{\bY}\,.
\end{eqnarray*}

Then the normal-gamma model 

\begin{equation*}
\tilde{\bY}|\bbeta,\sigma^2 \sim N(A\bbeta,\sigma^{-2}I_{nr})\,, \qquad \sigma^{-2} \sim \mbox{Gamma}(a_{\sigma},b_{\sigma})
\end{equation*}

\noindent is equivalent to

\begin{equation*}
\hat{\bbeta}|\bbeta,\sigma^2 \sim N(\bbeta,\sigma^2(A^{T}A)^{-1})\,, \qquad \sigma^{-2} \sim \mbox{Gamma}(a'_{\sigma},b'_{\sigma})\,.
\end{equation*}

\noindent The likelihood depends on the simulator data only through $\hat{\bbeta}$; therefore integrating out $\bbeta$ with respect to its prior distribution~(\ref{appbeta}) gives

\begin{eqnarray*}
 \pi ( \sigma^{-2},\btau,\btheta|\tilde{\bY}) &\propto& |(\sigma^{-2} A^{T}A)^{-1}+C|^{-\frac{1}{2}}\times\\\nonumber
&&\quad\exp\{ -\frac{1}{2}\hat{\bbeta}^{T}([\sigma^{-2} A^{T}A)^{-1}]+C]
)^{-1} \hat{\bbeta} \}\times\\\nonumber
&& (\sigma^{-2})^{a'_{\sigma}-1}\exp ^{-\sigma^{-2}/b'_{\sigma}}\times \prod_{i=1}^{p}\tau_{i}^{a_{\tau}-1}\exp ^{-\tau_{i}/b_{\tau}}\times \\\nonumber
&&\prod_{i=1}^{p}\left\{\prod_{k=1}^{q_{1}} \theta_{ik}^{a_{\theta}-1}(1-\theta_{ik}) ^{b_{\theta}-1}\right\}.
\end{eqnarray*}

This posterior distribution is explored via MCMC using standard metropolis updates. Conditional on the hyper-parameters, the posterior distribution of $\beta_k(\bx)$ is a Gaussian process of fairly standard form \cite{rougier2008}. Samples from the unconditional posterior for $\beta_k(\bx)$ can be obtained by substituting samples from the MCMC chain for $\sigma^{-2},\btau,\btheta|\tilde{\bY}$.

The MCMC simulation is hampered by the inversion of the matrix 

$$[(\sigma^{-2} A^{T}A)^{-1}+C]\,,$$ 

\noindent which is of size $np\times np$. However, only part of the matrix is updated at each step of the MCMC, therefore the matrix inversion is carried out once at the start of the chain and then updated via the method described in Appendix~\ref{secwood}. 


\section{Reduction of Computational Burden}
\label{secwood}

Inversion of the $np \times np$ matrix $[(\sigma^{-2} A^{T}A)^{-1}+C]$, see Appendix~\ref{itpsappendix}, is computationally intensive when using a non-orthogonal basis such as a thin plate spline. To overcome this problem and make the MCMC updates feasible, we note that any update of the parameters $\btau$, $\btheta$, or the nugget only change an $n \times n$ sub-matrix of the covariance matrix $C$. This sub-matrix depends on the parameter being updated, and hence care is required in locating the correct segment.  Once the inverse of the whole matrix has been performed for the update of $\sigma^{-2}$ the proceeding inverses can be computed using the Woodbury formula.

\begin{equation*}
(D+PQ)^{-1}= D^{-1}-D^{-1}P(I+QD^{-1}P)^{-1}QD^{-1},
\end{equation*}

\noindent where $D$, $P$ and $Q$ all denote matrices of the correct size, and $I$ is an identity matrix. For our problem, $D = [(\sigma^{-2} A^{T}A)^{-1}+C]$ and is size  $np \times np$, $P$ is size $np\times n$ and $Q$ is size $n \times np$.  In order to make use of this result, we need to find a $PQ$ equal to the difference in the $D$ matrix resulting from the update.  

Let 

\begin{eqnarray*}
P = \begin{pmatrix} \bz \\P_{1}\\ \bz \end{pmatrix}, \quad Q = \begin{pmatrix} \bz &Q_{1}& \bz \end{pmatrix}\,
\end{eqnarray*}

\noindent where each $\bz$ denotes a matrix of the correct dimension to position the change in the $D$ matrix for the current update.  Note that in the first updates, the initial $\bz$ matrix will be of size $0 \times 0$ and in the last updates the final $\bz$ matrix will be of size $0\times 0$.  Then 

\begin{eqnarray*}
PQ = \begin{pmatrix} \bz&\bz&\bz\\\bz& P_{1}Q_{1} &\bz \\ \bz&\bz&\bz\end{pmatrix}\,.
\end{eqnarray*}

We obtain $P_{1}Q_{1}$ by performing LU decomposition on the difference between the current $D$ matrix and the proposed $D$ matrix, a simple $np \times np$ subtraction. Using block notation and remembering that $D$ and therefore $D^{-1}$ are symmetric, let

\begin{eqnarray*}
D^{-1} = \begin{pmatrix} D_{11}^{-1} & D_{12}^{-1} & D_{13}^{-1} \\  D^{-T}_{12} & D^{-1}_{22}  & D^{-1}_{23} \\ D^{-T}_{13} & D_{23}^{-T} & D_{33}^{-1}
\end{pmatrix}\,.
\end{eqnarray*}

\noindent Then $QD^{-1}P$ can be replaced by $Q_{1}D_{22}^{-1}P_{1}$, and inversion of $(I+QD^{-1}P)^{-1}$ by inversion of $X = (I+ Q_{1}D_{22}^{-1}P_{1})^{-1}$, an $n \times n$ matrix.

Finally simple block multiplication of the defined matrices, taking account of symmetry, results in

\begin{eqnarray*}
(D+PQ)^{-1} & = & D^{-1}- \begin{pmatrix} D_{11}^{-1} & D_{12}^{-1} & D_{13}^{-1} \\  D^{-T}_{12} & D^{-1}_{22}  & D^{-1}_{23} \\ D^{-T}_{13} & D_{23}^{-T} & D_{33}^{-1}
\end{pmatrix}
\begin{pmatrix} \bz&\bz&\bz\\\bz& P_{1}X Q_{1} &\bz \\ \bz&\bz&\bz\end{pmatrix}\times \\ 
& &\hspace*{2cm} \begin{pmatrix} D_{11}^{-1} & D_{12}^{-1} & D_{13}^{-1} \\  D^{-T}_{12} & D^{-1}_{22}  & D^{-1}_{23} \\ D^{-T}_{13} & D_{23}^{-T} & D_{33}^{-1}
\end{pmatrix}\\
&=&D^{-1} -  \begin{pmatrix} D_{12}^{-1}P_1XQ_1D_{12}^{-T} & D_{12}^{-1}P_1XQ_1D_{22}^{-1} & D_{12}^{-1}P_1XQ_1D_{23}^{-1}\\
 D_{22}^{-1}P_1XQ_1D_{12}^{-T} & D_{22}^{-1}P_1XQ_1D_{22}^{-1}  & D_{22}^{-1}P_1XQ_1D_{23}^{-1} \\ 
 D_{23}^{-T}P_1XQ_1D_{12}^{-T} & D_{23}^{-T}P_1XQ_1D_{22}^{-1} & D_{23}^{-T}P_1XQ_1D_{23}^{-1} \end{pmatrix}.
\end{eqnarray*}

\noindent This can be computed efficiently using bespoke multiplication routines.  However, for large iteration cycles it is recommended that a full inversion be performed approximately once every 100 steps to ensure that small numerical errors do not appreciate. A similar derivation can be performed using the matrix determinant lemma \cite{harville}.


%
%
%
%

\bibliographystyle{siam}
\bibliography{dstl_mod}

\end{document}